\documentclass[pre,
 reprint,
 amsmath,amssymb,
 aps,
]{revtex4-2}
\def\onedot{$\mathsurround0pt\ldotp$}
\def\cddot{
  \mathbin{\vcenter{\baselineskip.67ex
    \hbox{\onedot}\hbox{\onedot}}%
  }}%
\def\cdddot#1{
  \mathbin{\vcenter{\baselineskip.67ex
    \hbox{\onedot}\hbox{\onedot}\hbox{\onedot}%
  }}%
}
\usepackage[caption=false]{subfig}
\usepackage[english]{babel}
\usepackage{graphicx}
\usepackage{dcolumn}
\usepackage{bm}
\usepackage{xcolor}

\definecolor{darkgreen}{rgb}{0,0.4,0.0}

\AtBeginDocument{\selectlanguage{english}}
\begin{document}

\preprint{APS/123-QED}

\title{Hydrodynamic Equations for Active Brownian Particles in the High Persistence Regime}

\author{Mart\'{\i}n Pinto-Goldberg, Rodrigo Soto}
\affiliation{Departamento de Física, FCFM, Universidad de Chile, Blanco Encalada 2008, Santiago, Chile}

\begin{abstract}

In the high persistence regime of non-inertial active Brownian particles (ABP), polarization becomes a relevant dynamical field. 
Based on a recently proposed kinetic description for ABP, we derive Navier--Stokes-like equations for the density and polarization fields in this regime.
Using the Chapman--Enskog method, all transport coefficients in the equations are obtained entirely in terms of the microscopic dynamics. 
A linear stability analysis of the homogeneous and isotropic state shows that the derived equations correctly describe the density instability associated to the motility induced phase separation. Numerical solutions of the equations in one spatial dimension show the need of an additional regularizing pressure term to saturate the system at high densities. With the inclusion of this term, the solutions illustrate in detail the clustering dynamics, with the formation of polarized regions at the interfaces, and the subsequent coarsening of domains, as well as particle accumulation in presence of gravity. Finally, the derived equations imply that, as an effect of the coupling with the polarization, damped density wave modes appear in the system which were verified with numerical simulations.

\end{abstract}

\maketitle

\section{Introduction}
In the framework of non-equilibrium statistical physics, the study of active matter~\cite{annurev:/content/journals/10.1146/annurev-conmatphys-070909-104101,RevModPhys.85.1143,shaebani2020computational} has gathered increasing interest in the last decades. From bacterial colonies \cite{Feng2017-lv} to schools of fish \cite{PhysRevE.109.064403}, the simple dynamics of individual active units can give rise to complex and rich collective behavior when interacting with one another. 
Self-propelled particles constitute a minimal model to such behavior, with the Vicsek model \cite{PhysRevLett.75.1226} providing a classic example of a flocking transition through alignment interactions. For simpler models, in the absence of alignment and even when the interactions are purely repulsive, processes such as clustering and phase separation can take place as a result of the persistent motion of the particles~\cite{PhysRevLett.108.235702,PhysRevLett.110.055701,PhysRevLett.110.055701}.

Active Brownian particles (ABP) \cite{Romanczuk2012-bu} constitute one of these models, in which spherical self-propelled particles of diameter $\sigma$ move persistently with velocity $V$ in a direction that changes diffusively over time at rate $D_r$. Despite its simplicity, the ABP model undergoes motility-induced phase separation (MIPS) \cite{Cates2013,annurev:/content/journals/10.1146/annurev-conmatphys-031214-014710} at high enough densities and persistence, the latter characterized by the active Péclet number $\text{Pe}=V/(\sigma D_r)$. The transition emerges from an interplay between volume exclusion and self-propulsion, where collisions reduce the effective velocity of particles in high density regions, producing further accumulation. The ABP model has motivated significant research due to its precision in describing some experimental systems in active matter, for instance, Janus colloids \cite{PhysRevLett.110.238301,BIALKE2015367} and Quincke Rollers \cite{PhysRevX.9.031043,Bricard2013-hy}. Numerically, ABP have been extensively studied, providing essential insight into the dynamics of phase separation ~\cite{PhysRevLett.110.055701,Gonnella2015-iv,PhysRevE.89.062301,PhysRevLett.108.235702,PhysRevE.103.022603,PhysRevResearch.2.023010,MacedoBiniossek_2018,C3SM52469H}. However, it still remains crucial to develop mathematical descriptions consistent with the observed dynamics, as a way to enable simpler analyses and make general predictions. 

Continuum descriptions prove to be a powerful tool to understand the emergent statistical properties of active systems \cite{RevModPhys.85.1143}. Many of these models have a phenomenological origin, stemming from symmetry arguments, numerical simulations, and experimental results \cite{annurev:/content/journals/10.1146/annurev-conmatphys-031214-014710,Wittkowski2014,Thampi2014,PhysRevLett.89.058101}. A remarkable example is the theory of flocking by Toner and Tu \cite{PhysRevLett.75.4326,PhysRevE.58.4828}, which proposes continuum equations of motion based solely on symmetry considerations. Another viable way to obtain macroscopic equations is to start from the microscopic dynamics and systematically build our way up with coarse-graining procedures \cite{SCHILLING20221}. Kinetic theory, as one of these methods, allows the study of a broad range of out-of-equilibrium phenomena with applications in many areas of physics \cite{kin}. In the realm of active matter, kinetic descriptions have been used to study particles with short-range polar order alignment \cite{PhysRevE.74.022101,PhysRevE.77.011920} as well as higher order interactions \cite{Boltz2024}. In analogy with the Toner and Tu theory, it has been shown that this framework allows to derive the hydrodynamic equations of flocking starting from first principles \cite{PhysRevE.77.011920}. 

Recently, a kinetic theory of overdamped ABP was developed, in which MIPS was derived starting from the microscopic interactions in the limit of infinite persistence~\cite{PhysRevLett.132.208301}. This theory redefines collisions due to encounters lasting a finite time. For this purpose, the interactions were characterized as instantaneous events producing effective displacements on the particles (this idea is further explained in Sec.~\ref{sec:Kinetic theory}). The kinetic formulation presents a foundation from which continuum equations can be obtained. Hydrodynamic theories for ABP usually focus on the density $\rho$ as the only relevant field \cite{Gonnella2015-iv,annurev:/content/journals/10.1146/annurev-conmatphys-031214-014710,Bialk2013,PhysRevLett.112.218304}. However, in the high persistence regime, $\text{Pe}\gg1$, the polarization $\mathbf{q}$ relaxes on times scales much larger compared to the microscopic dynamics. Furthermore, in wetting layers and at the interfaces of phase-separated profiles, orientational order emerges \cite{caprini_spontaneous_2020,hermann2020active,rojas2023wetting,perez2025two}, making the contribution of the polarization to the dynamics non-negligible. Experiments also show that synthetic microswimmers polarize against a gravity field, even in bulk-like regions \cite{Ginot_2018}, while numerical simulations reveal dissipative waves at low densities, far from the MIPS instability (see Sec.~\ref{sec:analysis}). For the reasons discussed above, in many situations, even in the absence of alignment interactions, a single-field theory and adiabatic slaving of the polarization $\mathbf{q}\sim\nabla[v(\rho)\rho]$~\cite{bialke_microscopic_2013}, is insufficient. This mirrors the role of the temperature in granular systems, where, although energy is not conserved in collisions \cite{garzo2019granular,brilliantov2004kinetic}, it is essential in giving a complete dynamical description of the flows.


In this work, starting from the aforementioned kinetic theory, we derive Navier--Stokes-like equations for the density and polarization fields to describe a system of ABP in the high persistence regime. With this aim, we employ the Chapman--Enskog method to the kinetic equation of ABP, which is a systematic approach to obtain equations that describe the system at different time scales.
The article is structured as follows. In Sec.~\ref{sec:Kinetic theory}, the kinetic equation describing the system is presented, along with the mathematical characterization of the effective collisions. Sec.~\ref{sec:slowmodes} presents the macroscopic variables relevant for a hydrodynamic description of the system. In Sec.~\ref{sec:ch-e}, the Chapman--Enskog method is developed, resulting in the hydrodynamic-like equations \eqref{eq:rhofull} and \eqref{eq:qfull}, which are the main contribution of this article. The equations obtained are examined in Sec.~\ref{sec:analysis} and a linear stability analysis shows the emergence of a density instability (MIPS). In Sec.~\ref{sec:gravity}, the hydrodynamics of the system are studied when a gravitational force is applied. Lastly, conclusions and perspectives are discussed in Sec.~\ref{sec:discussion}.

\section{Kinetic Theory}\label{sec:Kinetic theory}
\begin{figure}[h]
    \centering
    \includegraphics[width=\linewidth]{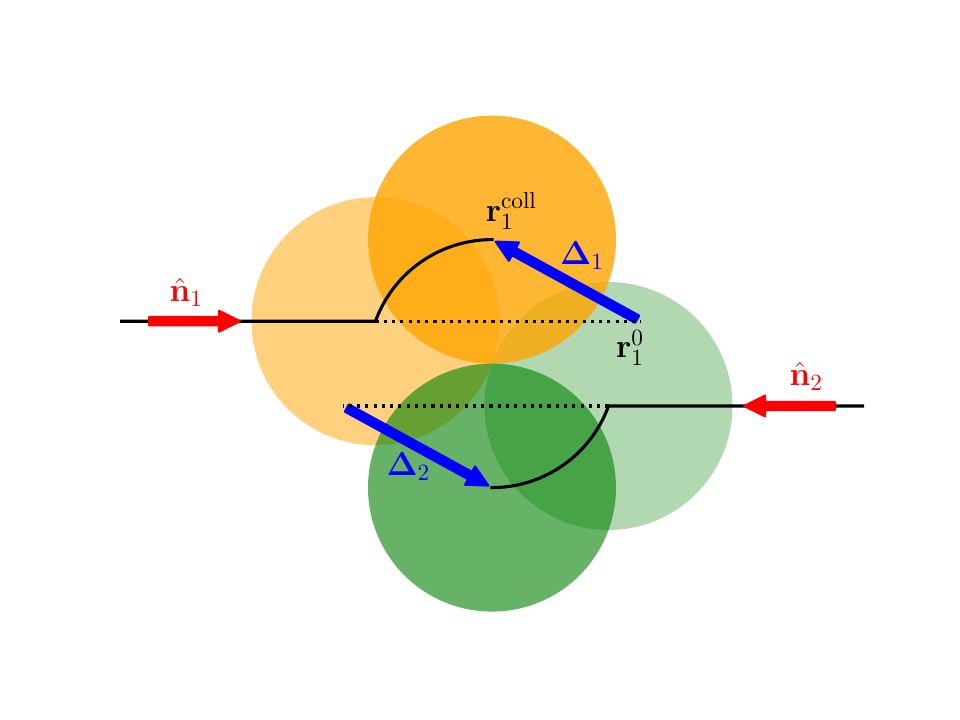}
    \caption{Collision schematic for ABP in the limit of infinite persistence. For simplicity we illustrate the case of $\mathbf{\hat{n}}_2=-\mathbf{\hat{n}}_1$. Particle directors are shown in red. The solid black lines represent the trajectories of the particles during the collision, while the dotted lines are the movement of the particles in the absence of it. The end points of these trajectories are $\mathbf{r}_i^{\rm{coll}}$ and $\mathbf{r}_i^{0}$, respectively. The effective displacements $\mathbf{\Delta}_i$ caused by the interaction are in blue.}
    \label{fig:coll}
\end{figure}

In two dimensions, the orientation of the ABP can be described by a single angle $\theta_i$ and the equations of motion for the $i-$th particle are 
\begin{equation}\label{eq:eom}
    \dot{\mathbf{r}}_i = V\mathbf{\hat{n}}_i+\mathbf{F}_i,\quad \dot{\theta}_i = \sqrt{2D_r}\xi_i(t),
\end{equation}
where $\mathbf{\hat{n}}_i = (\text{cos}\:\theta_i,\text{sin}\:\theta_i)$, $\xi_i(t)$ are uncorrelated white noises, and $\mathbf{F}_i$ imposes excluded volume. A kinetic theory description for ABP has been written following the effective collision theory presented in Ref.~\cite{PhysRevLett.132.208301}. In the present article, where the system is studied in two dimensions, the kinetic equation for the distribution function $f(\mathbf{r}_1,\mathbf{\hat n_1},t)$ reads
 \begin{equation}\label{eq:kin}
     \frac{\partial f}{\partial t} + V\mathbf{\hat{n}_1}\cdot\frac{\partial f}{\partial\mathbf{r}_1} = D_r\frac{\partial^2 f}{\partial \theta_1^2}+ J[f,f].
 \end{equation}
The first three terms account for the free streaming of particles and rotational diffusion, while the collision term is $J[f,f]$. As it is usual in kinetic theory, collisional terms consist of a gain and a loss term, which respectively increase and decrease the number of particles in a region of phase space. The case for active Brownian particles is no different, but it is more subtle. As two ABP come into contact, steric interactions prevent them from continuing their free flight, causing them to slide around each other until they can detach. In the high persistence regime, the directors will remain approximately constant during the time the particles are in contact $t_\text{coll}$, which is of order $\sigma/V$. Under this approximation, the interaction has no effect on the particle directors, but it does on their positions. A collision is then defined by considering the positions of the particles at the moment they enter in contact and at the moment of detachment. In Fig.~\ref{fig:coll}, two trajectories are depicted; the dotted black lines represent what the trajectories of the particles would be in the absence of collision, ending at $\mathbf{r}_i^{0}$, while the solid lines are the actual trajectories caused by the interaction, ending at $\mathbf{r}_i^{\rm{coll}}$.
Then, the collision can be thought of as an instantaneous event, where particles are  displaced by $\mathbf{\Delta}_i=\mathbf{r}_i^{\rm{coll}}-\mathbf{r}_i^{0}$ at the moment of contact. This displacement is such that after the particles are displaced, they evolve freely during a time $t_\text{coll}$ and end up at $\mathbf{r}_i^{\rm{coll}}$. The full calculations for the displacements $\mathbf{\Delta}_i = \mathbf{\Delta}_i(\mathbf{\hat{n}}_1,\mathbf{\hat{n}}_2,\hat{\bm{\sigma}})$ and the characterization in two dimensions is presented in Appendix~\ref{app:displ}

Then, the collision term is formulated as follows~\cite{PhysRevLett.132.208301}
\begin{multline}\label{eq:J}
    J[f,f] = \int f(\mathbf{r}'_1,\mathbf{\hat{n}}_1) f(\mathbf{r}'_1+\sigma\hat{\bm{\sigma}},\mathbf{\hat{n}}_2)
    |V\sigma (\mathbf{\hat{n}}_2-\mathbf{\hat{n}}_1)\cdot\hat{\bm{\sigma}}|\\\times\Theta[-(\mathbf{\hat{n}}_2-\mathbf{\hat{n}}_1)\cdot\hat{\bm{\sigma}}][\delta(\mathbf{r}_1-\mathbf{r}_1^\prime-\mathbf{\Delta}_1) -\delta(\mathbf{r}_1-\mathbf{r}^\prime_1)]d\mathbf{r}_1^\prime d\mathbf{\hat{n}}_2d\hat{\bm{\sigma}}.
\end{multline}
Here, $\hat{\bm{\sigma}}$ is the unit vector pointing from particle 1 to particle 2 when they meet and $\Theta$ is the Heaviside function which only selects collisional trajectories. The subtraction of the Dirac deltas represents the gain and loss terms. They can be interpreted as a particle lost at $\mathbf{r}_1$ at the moment the particles come into contact and gained at position $\mathbf{r}_1+\mathbf{\Delta}_1$. In this article, we are considering the case of ABP gases, in which case we can neglect the effect of pair correlations and consider a Boltzmann-like collision operator.

By noting that the subtraction of Dirac deltas can be written as $\delta(\mathbf{r}-\mathbf{r}_a)-\delta(\mathbf{r}-\mathbf{r}_b)=-\nabla_\alpha\int_{\mathbf{r}_b}^{\mathbf{r}_a}\delta(\mathbf{r-s})ds_\alpha$, the collisional term can be written as the divergence of a vector field, $J[f,f] = -\nabla_\alpha G_\alpha$, with
\begin{multline} \label{eq:JfnG}
G_\alpha(\mathbf{r}_1,\mathbf{\hat{n}}_1,t) 
    = \int f(\mathbf{r}'_1,\mathbf{\hat{n}}_1)f(\mathbf{r}'_1+\sigma\hat{\bm{\sigma}},\mathbf{\hat{n}}_2)\\
\times    |V\sigma (\mathbf{\hat{n}}_2-\mathbf{\hat{n}}_1)\cdot\hat{\bm{\sigma}}| \Theta[-(\mathbf{\hat{n}}_2-\mathbf{\hat{n}}_1)\cdot\hat{\bm{\sigma}}]\\
    \times\int_{\mathbf{r}'_1}^{\mathbf{r}'_1+\mathbf{\Delta}_1}\delta(\mathbf{r}-\mathbf{s})ds_\alpha d\mathbf{r}_1^\prime d\mathbf{r}_2^\prime d\mathbf{\hat{n}}_2d\hat{\bm{\sigma}}.
\end{multline}
Here, $\mathbf{G}$ represents the instantaneous mass flux due to particle teleportation at collisions, which is transmitted along the line connecting each pair of particles. This is analogous to the collisional transfer of momentum and energy for a hard sphere gas, which occurs instantaneously at the moment of collisions~\cite{irving1950statistical,kin}.

\section{Slow modes}\label{sec:slowmodes}
For ABP, the relevant observables can be defined as moments of the distribution function over the director $\mathbf{\hat{n}}_1$. In the high persistence regime, with $\rm{Pe}\gg1$, particles undergo several collisions before their orientation changes. Therefore, besides the particle density,
\begin{equation}\label{eq:rhoA}
    \rho(\mathbf{r}_1,t)\equiv \int f(\mathbf{r}_1,\mathbf{\hat{n}_1},t)d\mathbf{\hat{n}}_1,
\end{equation}
the mean particle orientation also becomes a relevant field for a hydrodynamic description of the system. Thus, we will study the polarization $\mathbf{q}$, defined as the orientation density
\begin{equation}
    \mathbf{q}(\mathbf{r}_1,t)\equiv\int f(\mathbf{r}_1,\mathbf{\hat{n}}_1,t)\mathbf{\hat{n}}_1d\mathbf{\hat{n}}_1 = \rho \mathbf{P}(\mathbf{r}_1,t),
\end{equation}
where $\mathbf{P}$ is the mean orientation. 
Integrating the kinetic equation (\ref{eq:kin}) with respect to $\mathbf{\hat{n}}_1$ yields the conservation equation for the density of particles 
\begin{equation}\label{eq:consrho}
    \frac{\partial \rho}{\partial t} = -\nabla \cdot\mathbf{J},
\end{equation}
where the particle flux is given by 
\begin{equation}\label{eq:fluxrho}      
\mathbf{J}(\mathbf{r}_1,t)=V\mathbf{q}(\mathbf{r}_1,t)+\int \mathbf{G}(\mathbf{r}_1,\mathbf{\hat{n}}_1,t)d\mathbf{\hat{n}}_1.
\end{equation}
Here, the second term represents the modification of the flux of particles due to collisions. Similarly, by multiplying Eq.~(\ref{eq:kin}) by $\mathbf{\hat{n}}_1$ and integrating over $\mathbf{\hat{n}}_1$ gives an equation for the polarization
\begin{equation}\label{eq:consq}
    \frac{\partial q_\nu}{\partial t} = - D_rq_\nu - \nabla_\alpha Q_{\alpha\nu},
\end{equation}
where $Q_{\alpha\nu}$ represents the components of the polarization flux tensor
\begin{multline}
    Q_{\alpha\nu}(\mathbf{r}_1,t)=V\int f(\mathbf{r_1},\mathbf{\hat{n}}_1,t)n_{1\alpha}n_{1\nu}d\mathbf{\hat{n}}_1\\+ \int G_\alpha(\mathbf{r}_1,\mathbf{\hat{n}}_1,t)n_{1\nu}d\mathbf{\hat{n}}_1.
\end{multline}
Once more, the second term is an additional flux that arises from collisions.

As it was advanced, Eq.~(\ref{eq:consq}) shows that the polarization is not stricly conserved, but when $D_r$ is small, then it evolves slowly and becomes a relevant field for describing the system. The conservation equations are not closed, since they depend on $\mathbf{J}$ and $Q_{\alpha\nu}$ which are unknown. In order to close the equations and obtain expressions for the fluxes in terms of the fields, the Chapman--Enskog method is applied.

\section{Chapman--Enskog solution}\label{sec:ch-e}
    The Chapman--Enskog method allows to obtain hydrodynamic equations in increasing orders of the inhomogeneities of the system. In the hydrodynamic regime of the system it should be possible to find functional forms of the distribution function that are completely described by the fields, which in turn follow the conservation equations (\ref{eq:consrho}) and (\ref{eq:consq}). To find these solutions, it is assumed that the kinetic equation admits normal solutions, meaning that the spatio-temporal dependence of $f$ is \textit{enslaved} to $\rho$ and $\mathbf{q}$~\cite{cercignani,chapman1990mathematical}
    \begin{equation}\label{eq:normal}
f(\mathbf{r}_1,\mathbf{\hat{n}}_1,t)=f(\mathbf{\hat{n}}_1;\rho(\mathbf{r}_1,t),\mathbf{q}(\mathbf{r}_1,t)).
\end{equation}
The assumption that, although not strictly conserved, $\mathbf{q}$ is a relevant field and that the distribution function can be enslaved to it, is analogous to the case of rapid granular flows. In that case, mass and momentum are conserved at collisions, but energy is not. Nevertheless, to derive the coarse grained equations, it is assumed that the distribution function can be slaved to the density, momentum, and granular temperature fields, with the latter being not conserved. The resulting equations accurately describe the dynamics of granular flows~\cite{brilliantov2004kinetic,garzo2019granular}.

When substituting (\ref{eq:normal}) into the kinetic equation, the time derivative can be written as 
\begin{equation}\label{eq:fieldschainrule}
    \frac{\partial f}{\partial t} = \frac{\partial f}{\partial \rho}\frac{\partial \rho}{\partial t}+\frac{\partial f}{\partial \mathbf{q}}\cdot\frac{\partial \mathbf{q}}{\partial t}.
\end{equation}
Then, a formal small parameter $\epsilon$ is introduced that multiplies the spatial gradients, meaning that the spatial inhomogeneites are small and the system is close to a state of equilibrium.
Later, $f$ is expanded in powers of $\epsilon$ 
    \begin{equation}\label{eq:fexpansion}
    f=f^{(0)} + \epsilon f^{(1)} + \epsilon^2 f^{(2)}+\dots.
\end{equation}
As this expansion increases the number of possible solutions, in the Chapman--Enskog method it is conventional to introduce the constraint 
\begin{equation}\label{eq:normalization}
    \int f^{(i)} \begin{pmatrix}1\\\mathbf{\hat{n}}_1\end{pmatrix} d\mathbf{\hat{n}}_1=\delta_{0i}\begin{pmatrix}
        \rho\\\mathbf{q}
    \end{pmatrix},
\end{equation}
i.e., it is demanded that the term of order $\epsilon^0$ reproduces exactly the hydrodynamic fields.

Lastly, time scale separation is introduced through several time variables, $t_0=t$, $t_1=\epsilon t$, $t_2 = \epsilon^2 t$, and so forth. With these variables, the dynamics on the specific time scales are reflected. For example, whenever $t\ll\epsilon^{-1}$, $t_1$ will be small and, at this time scale, the distribution function will depend only on $t_0$. Furthermore, when $t\gg\epsilon^{-1}$, $t_1$ will be very large and the system should have reached a stationary state on this time scale, implying that the distribution function depends on $t_2$. For this time scale separation to take place, it must be imposed that the distribution function is regular for $t_i\rightarrow0$ and $t_i\rightarrow\infty$, for all time scales. Then, the distribution function can be written in the form $f = f(\mathbf{\hat{n}}_1;\rho(\mathbf{r}_1,t_0,t_1,\dots),\mathbf{q}(\mathbf{r}_1,t_0,t_1,\dots))$ and the time derivative is expressed using the chain rule
\begin{equation}\label{eq:timeexpansion}
    \frac{\partial }{\partial t} = \frac{\partial }{\partial t_0}+\epsilon\frac{\partial }{\partial t_1}+\epsilon^2\frac{\partial }{\partial t_2}+\dots .
\end{equation}
In order to determine the hierarchy of equations, it also becomes necessary to expand the collision operator in gradients. It can be seen from Eq.~(\ref{eq:JfnG}) that the zeroth order term vanishes, meaning that the expansion has the form
\begin{equation}\label{eq:Jexpansion}
    J = \epsilon J^{(1)}+\epsilon^2 J^{(2)}+\dots .
\end{equation}

Introducing the expansions (\ref{eq:fexpansion}), (\ref{eq:timeexpansion}), and  (\ref{eq:Jexpansion}) into the kinetic equation (\ref{eq:kin}) yields a hierarchy of equations that can be solved systematically.

\subsection{Zeroth order equation}
At order $\epsilon^0$, the kinetic equation reads
\begin{equation}\label{eq:kin0}
\frac{\partial f^{(0)}}{\partial t_0}=D_r \frac{\partial^2 f^{(0)}}{\partial \theta_1^2}.
\end{equation}
In this equation, and for the following orders, the time derivatives with respect to each time scale must be understood in the sense of Eq.~(\ref{eq:fieldschainrule}).
Then, by taking moments of this equation, that is, multiplying by 1 and $\mathbf{\hat{n}}_1$ and integrating over $\mathbf{\hat{n}}_1$, it is found for the density 
\begin{equation}\label{eq:rho0}
\frac{\partial \rho}{\partial t_0}=0,
\end{equation}
and for the polarization
\begin{equation}\label{eq:q0}
\frac{\partial \mathbf{q}}{\partial t_0}=-D_r\mathbf{q}.
\end{equation}
These equations imply that while the density does not evolve in the fast time scale, the polarization does. Nonetheless, when $D_r$ is small, the decay of $\mathbf{q}$ is slow. By substituting the time derivatives of the fields back into Eq.~(\ref{eq:kin0}) we obtain a closed equation for $f^{(0)}$
\begin{equation}\label{eq:kin00}
    \mathcal{L}f^{(0)}=0,
\end{equation}
where the linear operator is  $\mathcal{L}=\mathbf{q}\cdot\partial/\partial\mathbf{q}+\partial^2/\partial\theta^2_1$. The kernel of the operator is a Fourier series of the form
\begin{equation}\label{eq:Lkernel}
    f^{(0)}=\sum_{nk}a_{nk}(\rho) q_x^{k} q_y^{n^2-k}e^{in\theta_1},
\end{equation}
with $n\in \mathbb{Z}$ and $k\in\{0,1,\dots,n^2\}$ in order to avoid singularities.
The coefficients for $n=0,\pm1$ are fixed by Eq.~(\ref{eq:normalization}). On the other hand, there is no apparent mathematical condition to fix the rest of the coefficients of the series. In order to continue to the next order, the solution (\ref{eq:Lkernel}) is truncated up to the terms which are linear in $\mathbf{q}$. As it will be shown, the polarization is small, meaning that contributions from the next orders, which scale as $\mathbf{q}^{n^2}$ are negligible. In tensorial notation the solution can be written as $A_0+\mathbf{A_1}\cdot\mathbf{\hat{n}}_1+\mathbb{A}_2\cddot\mathbf{\hat{n}_1\hat{n}_1}+\dots$ and for the two forms to be equivalent, the $A$-tensors must vanish if any two indices are contracted. In this way, the $i-$th term of the series is independent of the rest. With these assumptions the solution to the zeroth order equation is  
\begin{equation}\label{eq:f0}
    f^{(0)} = \frac{\rho}{2\pi} + \frac{\mathbf{q}\cdot\mathbf{\hat{n}}_1}{\pi}.
\end{equation}
This function is the analogous to a local Maxwellian distribution in hard sphere gases for ABP. This means that in the following orders, the resulting equations will be analogous to the Euler and Navier--Stokes equations, but now for the density and polarization fields.

\subsection{First order equation} \label{sec.1storder}
In the next order, $\epsilon^1$, the kinetic equation is
\begin{multline}\label{eq:kin1}
    \frac{\partial f^{(1)}}{\partial t_0} + \frac{\partial f^{(0)}}{\partial t_1}+ V\mathbf{\hat{n}}_1\cdot \nabla f^{(0)} = D_r \frac{\partial^2 f^{(1)}}{\partial\theta^2_1} + J^{(1)}[f^{(0)},f^{(0)}],
\end{multline}
where $J^{(1)}$ is the first term of the collisional integral expansion in gradients
\begin{multline}\label{eq:op1}
    J^{(1)}[f,g] = - \nabla_\alpha\int \Delta_{1\alpha}f_1g_2|V\sigma(\mathbf{\hat{n}}_2-\mathbf{\hat{n}}_1)\cdot\hat{\bm{\sigma}}|\\\times\Theta[-(\mathbf{\hat{n}}_2-\mathbf{\hat{n}}_1)\cdot\hat{\bm{\sigma}}]d\mathbf{\hat{n}}_2 d\hat{\bm{\sigma}},
\end{multline}
with $f_{1}=f(\mathbf{r}_1,\mathbf{\hat{n}}_{1})$ and $g_{2}=g(\mathbf{r}_1,\mathbf{\hat{n}}_2)$. Here, $\alpha$ denotes Cartesian coordinates and the Einstein summation convention is assumed.

When computing the moments of the kinetic equation at first order [Eq.~(\ref{eq:kin1})], the contribution of the first term vanishes due to the constraint of Eq.~(\ref{eq:normalization}). Furthermore, replacing $f^{(0)}$ with the solution (\ref{eq:f0}) gives
\begin{align}
\frac{\partial \rho}{\partial t_1} + V\nabla\cdot\mathbf{q}&=I^{(1)}[f^{(0)},f^{(0)}]\label{eq:rho1}, \\
\noindent\frac{\partial \mathbf{q}}{\partial t_1} + \frac{V}{2}\nabla\rho &= I^{(1)}_{\mathbf{\hat{n}}_1}[f^{(0)},f^{(0)}],\label{eq:q1}
\end{align}

where $I^{(i)}$ and $I^{(i)}_\mathbf{\hat{n}_1}$ are the moments of $J^{(i)}$
\begin{equation}
    \begin{pmatrix}
        I^{(i)}\\
        I_\mathbf{\hat{n}_1}^{(i)}
    \end{pmatrix}=\int J^{(i)}\begin{pmatrix}
        1\\\mathbf{\hat{n}_1}
    \end{pmatrix}d\mathbf{\hat{n}_1}.
\end{equation}
In Appendix \ref{app:integrals}, the term $J^{(1)}[f^{(0)},f^{(0)}]$ is calculated explicitly, as it is then straightforward to obtain its moments, which are
\begin{align}
    I^{(1)}[f^{(0)},f^{(0)}]&=0,\\
    I^{(1)}_\mathbf{\hat{n}_1\nu}[f^{(0)},f^{(0)}]&=\frac{V\pi\sigma^2}{8}[\nabla_\nu\rho^2-2\nabla_\alpha(q_\alpha q_\nu)].
\end{align}
After some simple algebra it is found that
\begin{align}
    \frac{\partial \rho}{\partial t_1} &=-V\nabla_\alpha q_\alpha,\label{eq:rho11}\\
    \frac{\partial q_\nu}{\partial t_1}&=-\frac{1}{2}\nabla_\nu[V_\text{eff}(\rho)\rho]-\frac{V}{\rho_l}\nabla_\alpha(q_\alpha q_\nu),\label{eq:q11}
\end{align}
where 
\begin{equation}
    V_\text{eff}(\rho)=V(1-\rho/\rho_l)
    \end{equation}
    is the effective velocity of convection for the polarization and $\rho_l=4/(\pi\sigma^2)$, with $l$ standing for the liquid phase.
    We note that this effective velocity appears directly from the analysis of the kinetic equation. This expression is consistent with that obtained from an heuristic analysis~\cite{PhysRevLett.132.208301} or a formal analysis of the self-diffusion process~\cite{10.1063/5.0255082} of the same kinetic equation.
    With these equations, analog to the Euler equations for classical fluids, it is already possible to predict the MIPS instability. Considering small harmonic perturbations to a homogeneous and isotropic reference state of density $\rho_0$ and vanishing polarization, the growth rates $\lambda$ of the different Fourier modes can be computed as a function of their wavevector $\mathbf{k}$. The calculations yield 
    \begin{equation}\label{eq:lambda1}
        \lambda_\rho(k) = -\frac{D_r}{2}+\frac{1}{2}\sqrt{D_r^2-2k^2V^2\left(1-2\rho_0/\rho_l\right)}
    \end{equation}
    for the density mode. For densities above the critical value $\rho_l/2$, the real part of $\lambda_\rho$ becomes positive and the mode grows rapidly. According to Eq.~(\ref{eq:lambda1}), above the instability, perturbations are unstable for all wavevectors, which make the system evolve to highly rough states. To suppress the instability at small wavelengths, it is necessary to obtain the equations at next order, to ensure there is saturation coming from diffusive terms. 
    
Using Eqs.\:(\ref{eq:rho11}) and (\ref{eq:q11}), and the explicit form of $f^{(0)}$, results in a closed equation for $f^{(1)}$ 
\begin{multline}\label{eq:kin111}
    \mathcal{L}f^{(1)} = -\frac{\nabla_\alpha}{2\pi D_r}[q_\alpha V_\text{eff}(\rho)-2q_\beta V_\text{eff}(\rho)n_{1\alpha}n_{1\beta}].
\end{multline}
The homogeneous solution of this equation has the form (\ref{eq:Lkernel}) and the normalization condition (\ref{eq:normalization}) imposes that the first two terms must vanish. This means that the leading order is $\mathbf{q}^4$ and, under the previous assumptions, is negligible. Then, only the particular solution is needed. Following the structure of the right-hand side of equation (\ref{eq:kin111}), the following ansatz is proposed
\begin{equation}
    f^{(1)}=A_0+A_{2\alpha\beta}n_{1\alpha} n_{1\beta},
\end{equation}
where the two terms must meet the condition $\rm{Tr}(A_{2\alpha\beta}) = -2A_0$ to satisfy normalization~\eqref{eq:normalization}. Replacing this ansatz in the equation gives a straightforward solution for $f^{(1)}$ 
\begin{equation}\label{eq:f1}
    f^{(1)}=\frac{1}{8\pi D_r}\nabla_\alpha\bigl[q_\alpha V_\text{eff}(\rho)-2q_\beta V_\text{eff}(\rho)n_{1\alpha}n_{1\beta}\bigr].
\end{equation}

\subsection{Hydrodynamic equations}
At the next order, the moments of the kinetic equation will give hydrodynamic-like equations with diffusion terms. The kinetic equation at order $\epsilon^2$ reads
\begin{multline}\label{eq:kin2}
    \frac{\partial f^{(0)}}{\partial t_2}+ \frac{\partial f^{(1)}}{\partial t_1}+ \frac{\partial f^{(2)}}{\partial t_0}+V\mathbf{\hat{n}}_1\cdot\nabla f^{(1)}=D_r\frac{\partial^2 f^{(2)}}{\partial \theta_1^2}\\ + J^{(1)}[f^{(0)},f^{(1)}]+J^{(1)}[f^{(1)},f^{(0)}]+J^{(2)}[f^{(0)},f^{(0)}],
\end{multline}
where $J^{(2)}$ is
\begin{multline}\label{eq:J2}
   J^{(2)}[f,g]= -\sigma\nabla_\alpha \int \Delta_{1\alpha}f_1 \hat{\sigma}_\beta\nabla_\beta g_2|V\sigma(\mathbf{\hat{n}}_2-\mathbf{\hat{n}}_1)\cdot\hat{\bm{\sigma}}|\\\times\Theta[-(\mathbf{\hat{n}}_2-\mathbf{\hat{n}}_1)\cdot\hat{\bm{\sigma}}] d\mathbf{\hat{n}}_2
   d\hat{\bm{\sigma}}+ \frac{1}{2}\nabla_\alpha \nabla_\beta \int\Delta_{1\alpha}\Delta_{1\beta} f_1g_2\\\times|V\sigma(\mathbf{\hat{n}}_2-\mathbf{\hat{n}}_1)\cdot{\hat{\bm{\sigma}}}|\Theta[-(\mathbf{\hat{n}}_2-\mathbf{\hat{n}}_1)\cdot\hat{\bm{\sigma}}]d\mathbf{\hat{n}}_2
   d\hat{\bm{\sigma}}.
\end{multline}
Taking the respective moments of the equations and using the solution for $f^{(1)}$ gives at this time scale 
\begin{align}
    \frac{\partial \rho}{\partial t_2} 
    &= I^{(1)}[f^{(0)},f^{(1)}] + I^{(1)}[f^{(1)},f^{(0)}] + I^{(2)}[f^{(0)},f^{(0)}] \label{eq:rhoB}, \\
    \frac{\partial \mathbf{q}}{\partial t_2}
    &= \frac{V}{16D_r} \nabla_\alpha^2 \bigl[ V_\text{eff}(\rho)\mathbf{q} \bigr]
    + I^{(1)}_{\mathbf{\hat{n}}_1}[f^{(0)},f^{(1)}]
    + I^{(1)}_{\mathbf{\hat{n}}_1}[f^{(1)},f^{(0)}] \notag \\
    &\quad + I^{(2)}_{\mathbf{\hat{n}}_1}[f^{(0)},f^{(0)}]. \label{eq:q}
\end{align}
The calculation of the moments of the collisional integral is straightforward but tedious, and it is left for Appendix~\ref{app:integrals} for the interested reader. The final results give 
\begin{equation}\label{eq:rho2}
    \frac{\partial \rho}{\partial t_2} = A_1\nabla^2\rho^2-A_2\nabla^2\mathbf{q}^2-3A_3\nabla_\alpha\nabla_\beta(q_\alpha q_\beta),
\end{equation}
\begin{multline}\label{eq:q2}
    \frac{\partial q_\nu}{\partial t_2}=A_4\nabla_\alpha^2[V_\text{eff}(\rho)q_\nu]-A_5\left\{2\nabla_\alpha[ \rho\overline{\nabla_\alpha(V_\text{eff}(\rho)q_\nu)} ]\right.\\-\nabla_\nu[ \rho\nabla_\alpha(V_\text{eff}(\rho)q_\alpha)
 ]\Bigr\}+A_3\left[\nabla_\alpha^2(\rho q_\nu)-2\nabla_\nu\nabla_\alpha(\rho q_\alpha)\right.\\\left.+\nabla_\alpha(q_\alpha\nabla_\nu\rho)+\nabla_\nu(q_\alpha\nabla_\alpha\rho)  \right]+A_6\nabla_\alpha(q_\nu\nabla_\alpha\rho),
\end{multline}
where the overline denotes a symmetrized tensor $\overline{D_{\alpha\beta}}=(D_{\alpha\beta}+D_{\beta\alpha})/2$, $A_i$ are coefficients associated to nonlinear diffusion processes
\begin{align*}
    &A_1 = \frac{2V\sigma^3C}{\pi}&A_2 &= \frac{2V\sigma^3}{3\pi}(1-C)&A_3&=\frac{4V\sigma^3}{9\pi}(3C-1)\\
    &A_4=\frac{V}{16D_r}&A_5&=\frac{V\pi\sigma^2}{64D_r}&A_6&=\frac{4V\sigma^3}{9\pi}(3C+1)
\end{align*}
and $C\approx0.916$ is the Catalan constant. 
To obtain the general equation for the fields, the dynamics at all orders are summed by replacing $t_n = \epsilon^n t$ and then $\epsilon$ is set to unity
\begin{widetext}
\begin{align}
\label{eq:rhofull}
    \frac{\partial \rho}{\partial t} =&-V\nabla_\alpha q_\alpha +A_1\nabla^2\rho^2-A_2\nabla^2\mathbf{q}^2-3A_3\nabla_\alpha\nabla_\beta(q_\alpha q_\beta),
\\
\frac{\partial q_\nu}{\partial t}=&-D_r q_\nu -\frac{1}{2}\nabla_\nu[V_\text{eff}(\rho)\rho]-\frac{V}{\rho_l}\nabla_\alpha(q_\alpha q_\nu)+A_4\nabla_\alpha^2[V_\text{eff}(\rho)q_\nu]
-A_5\left\{2\nabla_\alpha\left[ \rho\overline{\nabla_\alpha(V_\text{eff}(\rho)q_\nu)} \right]\right.\nonumber\\
&-\nabla_\nu\left[ \rho\nabla_\alpha(V_\text{eff}(\rho)q_\alpha)
 \right]\Big\}+A_3\left[\nabla_\alpha^2(\rho q_\nu)-2\nabla_\nu\nabla_\alpha(\rho q_\alpha)+\nabla_\alpha(q_\alpha\nabla_\nu\rho)+\nabla_\nu(q_\alpha\nabla_\alpha\rho)  \right]+A_6\nabla_\alpha(q_\nu\nabla_\alpha\rho).\label{eq:qfull}
\end{align}
These equations constitute the main result of this article. They describe the dynamics of a gas of ABP in the high persistence regime, where the density and polarization fields are the relevant fields. They are based on pure microscopic arguments, with no adjustable parameters. In the next section, we analyze them showing that, for example, they are able to describe the MIPS instability.
\end{widetext}

\section{Analysis}\label{sec:analysis}
\subsection{Preliminary remarks}
Having derived the hydrodynamic equations for $\rho$ and $\mathbf{q}$ from first principles, it is remarkable that a density dependent velocity $V_\text{eff}(\rho)$ appears in the convection term for the polarization. It does not appear in the equation for the density, contrary to what has been proposed using other theoretical approaches~\cite{Bialk2013, Cates2013}.

The diffusive terms in the equations are solely nonlinear, which stems directly from the absence of translational noise in the model, Eq.~(\ref{eq:eom}). This means that diffusion is achieved exclusively through the repulsive interactions between particles. These terms are indeed collisional because the coefficients in front of them are proportional to powers of $\sigma$, vanishing for point particles.

The polarization decays at a constant rate $D_r$ and the form of Eq.~(\ref{eq:qfull}) suggests that after  transients its value scales as $\nabla\rho$ and higher order gradients. In the spirit of the Chapman--Enskog procedure, this means that $\mathbf{q}$ is small, which justifies neglecting terms like $\mathbf{q}^4$ and higher powers. 

It is worth noting that terms alluding to a convective derivative $\mathbf{q}\cdot\nabla\mathbf{q}$ originate entirely from collisions, which can easily be seen by the presence of factors $\sigma$. In kinetic theory of gases \cite{kin}, these terms appear naturally from the free streaming terms in the kinetic equation. This absence can be explained by the lack of  $\mathbf{q}^2$ terms in the solution to the zeroth order equation (\ref{eq:Lkernel}).

\subsection{Linear Stability Analysis}
The hydrodynamic equations (\ref{eq:rhofull}) and (\ref{eq:qfull}) can be used to study the stability of the homogeneous state. Consider a homogeneous and isotropic reference state of density $\rho=\rho_0$ and polarization $\mathbf{q}=0$. Then, considering small perturbations of the form $\rho(\mathbf{r},t)=\rho_0+\delta\rho(t) e^{i\mathbf{k}\cdot\mathbf{r}}$ and $\mathbf{q}=\delta\mathbf{q}(t)e^{i\mathbf{k}\cdot\mathbf{r}}$ we linearize the equations, leading to
\begin{align}
\frac{\partial \delta\rho}{\partial t} =& -iVk_\alpha \delta q_{1\alpha}-2A_1\rho_0k_\alpha^2\delta\rho,\\
\frac{\partial \delta q_{\nu}}{\partial t} =& -D_r\delta q_{\nu}-\frac{ik_\nu}{2}V_\text{eff}(2\rho_0)\delta\rho\nonumber\\-&\left[A_3\rho_0+\frac{V_\text{eff}^2(\rho_0)}{16D_r}\right]k_\alpha^2\delta q_{\nu}+2A_3\rho_0k_\nu k_\alpha \delta q_{\alpha} .\label{eq:q1linear}
\end{align}
Next, we proceed by separating the transverse and longitudinal components of the polarization by writing $\delta\mathbf{q}=\delta q_{\parallel}\mathbf{\hat{k}}+\delta q_{\perp}\mathbf{\hat{k}}_\perp$, where $\mathbf{\hat{k}}_\perp$ is the unit vector perpendicular to the wave vector. Then, by projecting Eq.~(\ref{eq:q1linear}) perpendicular to and along $\mathbf{\hat{k}}$ we can decouple the equations to find a system of linear equations in the perturbations
\begin{equation}\label{eq:linearperturbation}
    \frac{\partial }{\partial t}\begin{pmatrix}
        \delta\rho\\\delta q_\parallel\\\delta q_\perp\end{pmatrix}
        =M(k;\rho_0) \begin{pmatrix}
        \delta\rho\\\delta q_\parallel\\\delta q_\perp\end{pmatrix},
\end{equation}
where the coupling matrix $M(k;\rho_0)$ is
\begin{widetext}  
\begin{equation}
        M(k;\rho_0)=\begin{pmatrix}
            -2A_1\rho_0k^2&-ikV&0\\
            -\frac{ik}{2}V_\text{eff}(2\rho_0)&-D_r+\left(A_3\rho_0-\frac{V_\text{eff}^2(\rho_0)}{16D_r}\right)k^2 &0\\
            0&0&-D_r-\left(A_3\rho_0+\frac{V_\text{eff}^2(\rho_0)}{16D_r}\right)k^2
        \end{pmatrix}.
\end{equation}
\end{widetext}

The first thing to notice is that the transverse component $q_\perp$ decouples from the other modes. Furthermore, the eigenvalue $\lambda_{q_\perp}$ associated to this component is always negative, meaning that this mode is stable for all wavevectors.

\begin{figure}
\centering
{
\includegraphics[width=.49\linewidth]
{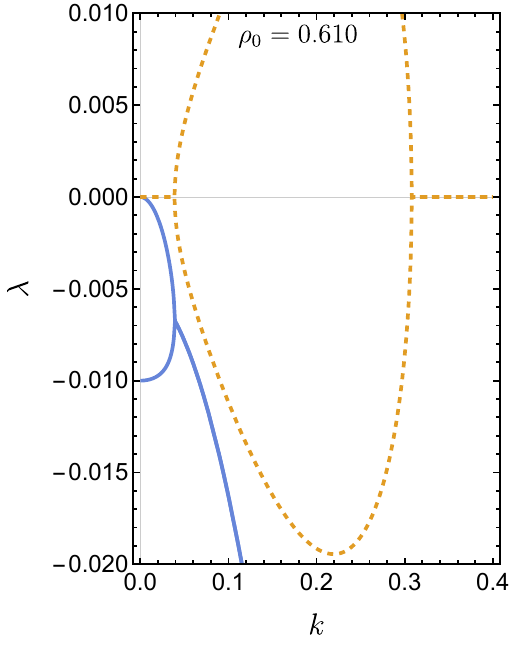}}
{\includegraphics[width=.49\linewidth]{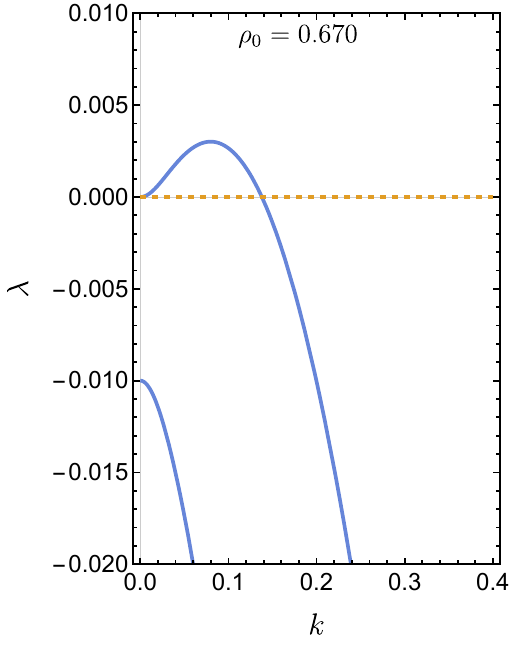}}

\caption{Eigenvalues of the coupled modes, $\rho$ and $q_\parallel$ as a function of the wavevector $k$ for $D_r=0.01$ and $\rho_0=0.61$ (left) and $\rho_0=0.67$ (right). The real (imaginary) parts are shown with solid blue (dashed orange) lines. Units have been chosen so that $V=\sigma=1$.}
\label{fig:lambdas}
\end{figure}

For the longitudinal part, which couples the density and parallel polarization fields, the phenomenology is  richer. Finding the eigenvalues of the coupled modes is straightforward, but the explicit expressions are not written down due to their length.
At vanishing wavevectors, one eigenvalue vanishes (the density mode) and the other is equal to $-D_r$ (the polarization mode). Figure~\ref{fig:lambdas} shows the eigenvalues as a function of the wavenumber $k$ for two densities, signaling that there is a critical value $\rho^*$, where the real part of the density mode  becomes positive. The polarization mode remains stable for all physically possible densities, $\rho<\rho_\text{max}=2/\sqrt{3}\sigma^2$.
The critical density at which the instability appears is given by $d^2\lambda_\rho/dk^2|_{k=0}=0$. This condition gives a simple expression for $\rho^*$ as a function of the persistence
\begin{equation}\label{eq:rhocrit}
    \rho^*=\frac{\rho_l/2}{1-16C/(\pi^2\text{Pe})}.
\end{equation}
This value agrees with the one found in Ref.\:\cite{PhysRevLett.132.208301} when doing a MIPS analysis for the reduction of the velocity due to collisions and a linear stability analysis of the kinetic equation. Also, in the limit $\text{Pe}\to\infty$, $\rho^*$ coincides with the critical density using the Euler-like equations of Sec.~\ref{sec.1storder}.


\begin{figure}
    \centering
    \includegraphics[width=\linewidth]{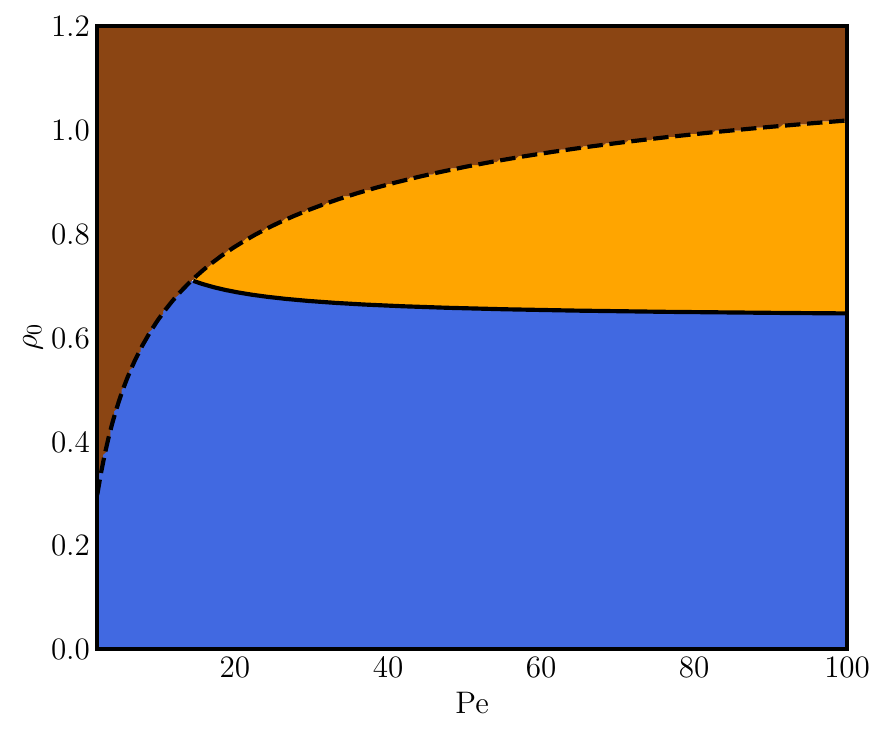}
    \caption{Phase diagram in the $\rm{Pe}-\rho_0$ space. Above the dashed line, the brown region shows where the theory loses validity. The solid black line shows the critical density: above (below) the solid black line the unstable (stable) region is shown in orange (blue). Units were fixed by $V=\sigma=1$.}
    \label{fig:phase}
\end{figure}

The analysis of the eigenvalues shows also that for each value of the system density, there is critical value of $D_r$ above which the density eigenvalue remains positive up to infinitely large wavevectors. This absence of saturation implies that the system would evolve to an extremely rough state, with wavelengths even smaller than the particle size, being therefore nonphysical.
This imposes a mathematical condition for the validity of the theory. By studying the behavior of $\lambda_\rho$ at large $k$, we found the critical value of $D_r$ as a function of the density to be

\begin{equation}\label{eq:Dr*}
    D_r^\text{c} = \frac{V_\text{eff}^2(\rho)}{16A_3\rho}.
\end{equation}
By inverting this relation, we can complete the phase diagram shown in Figure \ref{fig:phase}.

\subsection{Density-polarization oscillations}

Figure \ref{fig:lambdas} shows that for densities below the instability, there is a range in $k$ where the two modes merge and the eigenvalues become complex (for example, in Fig.~\ref{fig:lambdas}-left with $\rho_0=0.61$ and $D_r=0.01$, this happens for $0.05<k<0.3$).  These correspond to damped density-polarization waves. To verify this prediction, we performed simulations of ABP systems with periodic boundary conditions (details in the Appendix~\ref{app:simulations}). In those simulations, we measure in the steady state the intermediate scattering function
\begin{align}
F_\textbf{k}(t) = \frac{1}{N} \langle \rho_\textbf{k}(t)\rho_\textbf{k}^*(0)\rangle,
\end{align}
where $\rho_\textbf{k}(t)=\sum_{i=1}^N e^{i\textbf{k}\cdot\textbf{r}_i}$ is the Fourier transform of the instantaneous density, with $\mathbf{r}_i$ the particle positions. The temporal evolution of $F_\textbf{k}(t)$ gives information of the relevant dynamical modes in the system.

\begin{figure}[h]
    \centering
    \includegraphics[width=\linewidth]{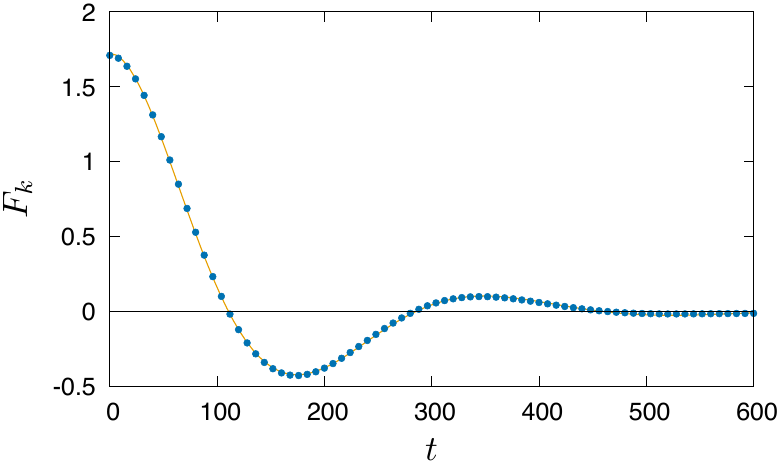}
    \caption{Intermediate scattering function obtained in a simulation of $N=6400$ ABP placed in square box of size 
    $L\approx178.9$ (average density $\rho=N/L^2=0.2$) for $k=2\pi/L$. The dots are the results of the simulations and the solid line is a fit to the function $F_\text{oscill}(t)=A e^{-t/\tau}\left[\cos\omega t+\sin\omega t/(\omega \tau) \right]$. Units have been chosen so that $V=\sigma=1$.}
    \label{fig:Fk-square}
\end{figure}

\begin{figure}[h]
\centering
\includegraphics[width=\linewidth]{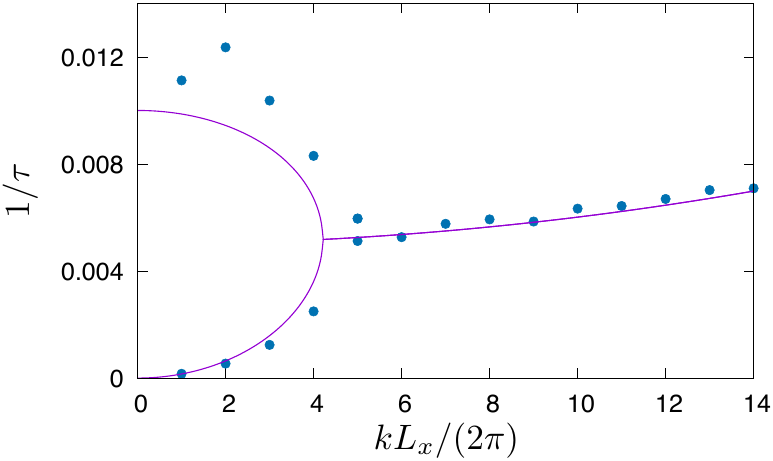}
\includegraphics[width=\linewidth]{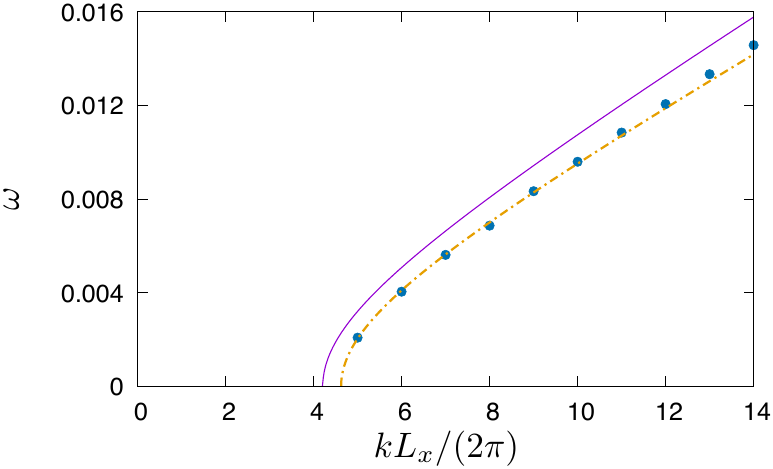}
\caption{Top: inverse values of the fitted relaxation times $\tau_{1,2}$ and $\tau$ as a function of the dimensionless wavenumber for simulations of $N=6000$ ABP in a rectangular box of lengths $L_x=3000$ and $L_y=10$ (average density $\rho=0.2$), and $\text{Pe}=100$. The dots are the results of the simulations and the solid line is the theoretical prediction. Bottom: fitted oscillation frequencies $\omega$. The dots are the results of the simulations, the solid line is the theoretical prediction, and the dashed line the fit to $\omega=C\sqrt{k^2-k_0^2}$, which gives $k_0=4.62\times 2\pi/L_x$. Units have been chosen so that $V=\sigma=1$.}
    \label{fig:Fk-tau-omega}
\end{figure}

First, we consider a system with $N=6400$ particles  placed in a square box of size $L\approx178.9\sigma$, such that the average density is $\rho=N\sigma^2/L^2=0.2$ and $\text{Pe}=100$, deep into the stable region. Figure~\ref{fig:Fk-square} shows $F_\textbf{k}(t)$ for $k=2\pi/L$, where it is clear the presence of damped oscillations. To obtain the transition to pure exponential decays for smaller wavevectors, we consider a system with $N=6000$ placed in a rectangular box of lengths $L_x=3000\sigma$ and $L_y=10\sigma$, with the same average density $\rho=0.2$ and Péclet number $\text{Pe}=100$. Here, the $F_\mathbf{k}(t)$ is measured for wavevectors along the long direction $\mathbf{k}=2\pi n/L_x \hat{\mathbf{x}}$ for $n=1,2,\dots,14$. In this case, depending on the valued of $k$, pure exponential decays or damped oscillations are obtained for $F_\mathbf{k}(t)$, which are fitted respectively to $F_\text{pure-exps}(t)= A\left(\tau_1e^{-t/\tau_1} - \tau_2e^{-t/\tau_2}\right)/(\tau_1-\tau_2)$ and $F_\text{oscill}(t)=A e^{-t/\tau}\left[\cos\omega t+\sin\omega t/(\omega \tau) \right]$, where it has been imposed that $dF(0)/dt=0$. Figure~\ref{fig:Fk-tau-omega}-top presents the fitted values of $\tau_{1,2}$ and $\tau$, while the bottom plot presents the obtained frequencies when the intermediate scattering functions present oscillations. The obtained results are consistent with the predictions of the hydrodynamic equations in that the two modes couple in the form of damped oscillations. The numerical agreement with the theory is very good. The wavenumber where the modes merge is predicted to be $k_0=4.23\times 2\pi/L_x$, while the simulations give $k_0=4.62\times 2\pi/L_x$. The deviations have their origin in that particle correlation cannot be fully discarded for this density. 

\subsection{Adiabatic elimination of the polarization}
Even though the equations~(\ref{eq:rhofull}) and (\ref{eq:qfull}) were derived for the long persistence regime, they can be analyzed for short persistence lengths. In this limit, the polarization relaxes quickly,
\begin{equation}\label{eq:qadiabatic}
    \mathbf{q}=-\frac{1}{2D_r}\nabla [V_\text{eff}(\rho)\rho]+\mathcal{O}(\nabla^3),
\end{equation}
meaning that it becomes enslaved to the density field. 
By substituting this expression into the density equation and keeping terms up to second order in gradients we obtain a diffusion equation,
\begin{equation}\label{eq:rhoadiabatic}
    \frac{\partial \rho}{\partial t} = \nabla\cdot[D_0(1-g\rho)\nabla\rho],
\end{equation}
where $D_0=V^2/(2D_r)$ is the bare diffusion coefficient for ABP and $g = (1-16C/\pi^2{\rm Pe})/(\rho_l/2)$. If the Chapman--Enskog method was applied only considering the density as a slow field, the same equation would have been obtained to this order in gradients. 

It is worth noting that Eq.~(\ref{eq:rhoadiabatic}) becomes unstable when $1-g\rho<0$, condition that gives the same critical density obtained in Eq.~(\ref{eq:rhocrit}). Note that above the critical density, all wavevectors are unstable. To obtain saturation for finite wavevectors, it would be necessary to apply the Chapman--Enskog method to higher orders and thus obtain Burnett and super-Burnett-like equations for ABP.

\subsection{Quasi one-dimensional system}\label{sec:analysis1d}
In a system with spatial symmetry along one dimension, the dependence of the fields can be restricted to a single spatial dimension, $\rho=\rho(x,t)$ and $\mathbf{q}=q(x,t)\mathbf{\hat{x}}$ and the equations (\ref{eq:rhofull}) and (\ref{eq:qfull}) can be somewhat simplified
\begin{align}\label{eq:rho1d}
    \frac{\partial \rho}{\partial t}=&-V\frac{\partial q}{\partial x}+\frac{\partial^2}{\partial x^2}\left[A_1\rho^2-(A_2+A_3)q^2\right],\\
\frac{\partial q}{\partial t}=&-D_rq-\frac{\partial}{\partial x}\left[\frac{1}{2}V_\text{eff}(\rho)\rho+\frac{V}{\rho_l}q^2\right]\nonumber\\
&+\frac{\partial^2}{\partial x^2}\left[A_4V_\text{eff}(\rho)q-A_6\rho q  \right]+(2A_6+A_7)\frac{\partial}{\partial x}\left(q\frac{\partial\rho}{\partial x}\right)\nonumber\\
&-A_5\frac{\partial}{\partial x}\left\{\rho\frac{\partial}{\partial x}[V_\text{eff}(\rho)q]\right\}.\label{eq:q1d}
\end{align}
These equations can be solved numerically, starting from a homogeneous state $\rho(x,0)=\rho_0$ and a small Gaussian perturbation for the polarization $q(x,0)=\xi(x)$. Fixing the units by $V=\sigma=1$, the equations are solved for $\rm{Pe}=80$. For values of $\rho_0$ below the critical density, the equations correctly simulate the system evolving to a homogeneous state. Above the critical value of the density there is indeed a density instability, but at long times it gives rise to diverging results: after the small wavelengths in the initial perturbation are suppressed, density peaks get narrower and grow past the value $\rho = \rho_l\approx 1.27$, where the effective velocity becomes negative, thus yielding an non-physical situation. The kinetic theory presented above was derived in the dilute regime, and while it predicts MIPS, it is not expected to describe what happens after the instability, particularly in the dense regime. Then, it can be anticipated that the hydrodynamic equations do not describe the phase separation accurately. 

\begin{figure}
    \centering
    \includegraphics[width=1.01\linewidth]{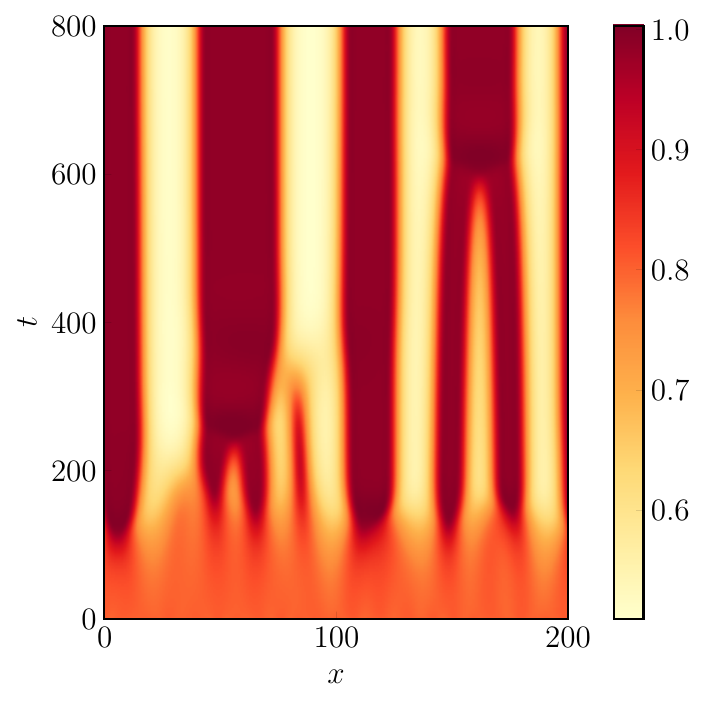}
    \caption{Spatiotemporal evolution of the density $\rho(x,t)$ for $D_r = 1/80$, starting from a homogeneous state $\rho(x,0)=0.8$ and $q(x,0)=\xi(x)$ a Gaussian white noise with zero mean and variance $0.01$. 
    Periodic boundary conditions were imposed and units were fixed by $V=\sigma=1$.}
    \label{fig:rho1d}
\end{figure}

\begin{figure}
    \centering
    \includegraphics[width=1.01\linewidth]{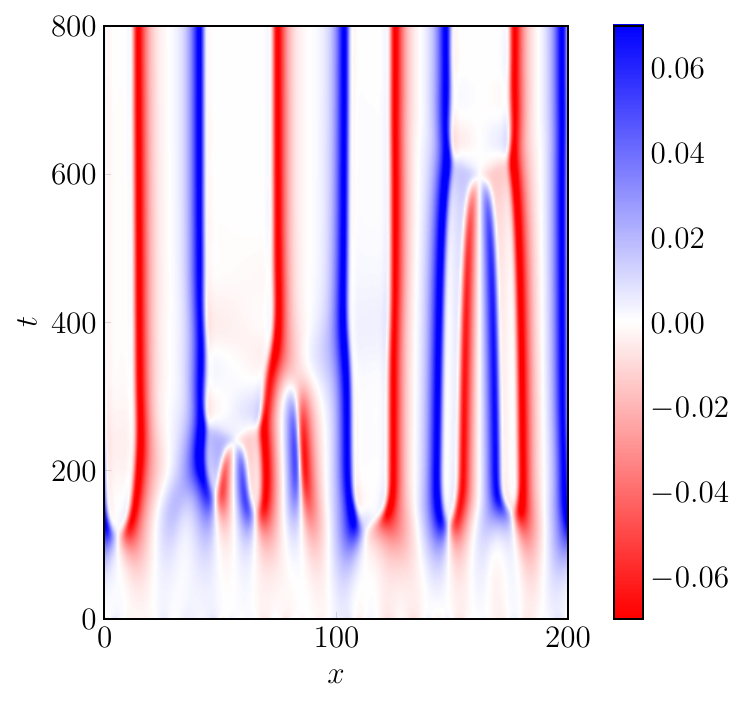}
    \caption{Spatiotemporal evolution of the polarization $q(x,t)$, with the same parameters as in Fig.~\ref{fig:rho1d}. The red (blue) regions indicate where the polarization is positive (negative), meaning that the particles are pointing right (left). The white regions have zero polarization.}
    \label{fig:q1d}
\end{figure}

In order to regularize the equations and avoid diverging solutions, we can take inspiration from the Euler equations for classical fluids, particularly from the velocity equation in absence of external forces
\begin{equation}\label{eq:euler}
    \frac{\partial (\rho u_\alpha)}{\partial t} = -\nabla_\beta(\rho u_\alpha u_\beta+ P_{\alpha\beta}),
\end{equation}
where $\rho$ is the mass density, $\mathbf{u}$ is the velocity field and $P_{\alpha\beta}$ is the pressure tensor. By looking at the form of (\ref{eq:q1d}) up to first derivatives, we can understand the second term on the right hand side as the gradient of pressure, $P = V_\text{eff}(\rho)\rho/2$. With this consideration, we propose to add a term to the pressure that becomes large close to the liquid density, effectively slowing mass transport and stopping the density to grow past non-physical values. The specific form we choose is 
\begin{equation}\label{eq:pressure}
P = \frac{1}{2}V_\text{eff}(\rho)
\rho+0.01V\rho\frac{ (\rho\sigma^2)^{14}}{1-\rho/\rho_l}, 
\end{equation}
which was chosen such that the dynamics are not modified for small densities, but diverges at $\rho_l$, before the effective velocity can become negative. Considering the additional term, the density saturates at a finite value, as shown in Figs.~\ref{fig:rho1d} and \ref{fig:q1d}.
In the unstable region of the phase diagram, it can be seen that the initial perturbation in the polarization causes clusters to form through spinodal decomposition. After the initial clusters reach the liquid density, they coarsen by coalescence, slowly moving closer until they merge (see movie in the Supplemental Material). Longer numerical solutions suggest that given enough time, the coarsening process should lead to macroscopic phase separation between the dense and dilute regions. In one dimension, we were able to compute the average domain size as $\ell(t)=L/N(t)$, where $N(t)$ is the number of domains as a function of time. For this purpose, we average over 4 solutions of the equations in a large box of $L = 2000\sigma$, which vary on the random initial condition, for times up to $t = 8\times10^5 \sigma/V$. 
\begin{figure}
    \centering
    \includegraphics[width=\linewidth]{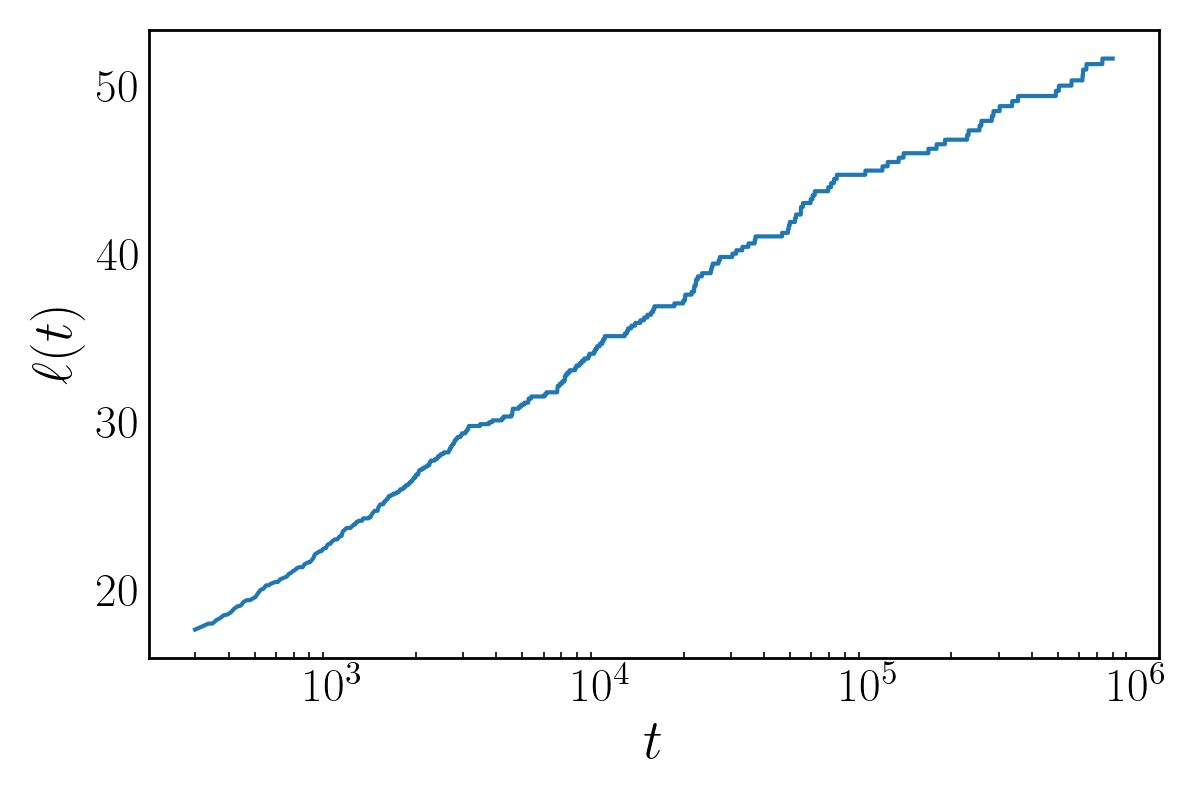}
    \caption{Average domain size as a function of time for $L=2000$ and $\text{Pe}=80$. Units were fixed by $V=\sigma=1$. }
    \label{fig:coarsening}
\end{figure}
Figure \ref{fig:coarsening} shows that the coarsening is logarithmic, $\ell(t)\sim \ln t$, which is consistent with the passive coarsening law for conservative dynamics of Hamiltonian systems (model B dynamics) in the absence of noise for one dimensional systems~\cite{langer1971theory,bray2002theory}.  

It should be noted that the values of the coexisting densities are  influenced by the choice of $P$. Thus, for better quantitative predictions, a more complete kinetic theory should be built, considering director reorientations during collisions for finite ${\rm Pe}$ and taking into account spatial correlations in the calculation of the hydrodynamic equations.

\section{ABP in presence of gravity}\label{sec:gravity}
When considering the system of active particles in a constant force field, as in the case of  gravity, the overdamped equations of motion for the positions become
\begin{equation}
    \dot{\mathbf{r}}_i = V\mathbf{\hat{n}}_i + \mathbf{V_g}+\mathbf{F}_i,
\end{equation}
where $\mathbf{V_g}=v_g\mathbf{\hat{g}}$, with $v_g$ being the magnitude of the sedimentation velocity due to the gravity and $\mathbf{\hat{g}}$ the unit vector pointing in the direction of it. It is important to note that when adding this constant velocity to the equations, the expressions for the displacements at particle collisions remain unchanged (see Appendix \ref{app:displ}). Thus, the only change to the conservation equations~(\ref{eq:consrho}) and (\ref{eq:consq}) are additional flux terms:
\begin{align}
\mathbf{J}(\mathbf{r}_1,t)=&V\mathbf{q}(\mathbf{r}_1,t)+\int \mathbf{G}(\mathbf{r}_1,\mathbf{\hat{n}}_1,t)d\mathbf{\hat{n}}_1+\mathbf{V_g}\rho,\\
Q_{\alpha\nu}(\mathbf{r}_1,t)=&V\int f(\mathbf{r_1},\mathbf{\hat{n}}_1,t)n_{1\alpha}n_{1\nu}d\mathbf{\hat{n}}_1\nonumber\\
&+ \int G_\alpha(\mathbf{r}_1,\mathbf{\hat{n}}_1,t)n_{1\nu}d\mathbf{\hat{n}}_1+V_{g\alpha}q_\beta.
\end{align}
Then, by construction, in the Chapman--Enskog solution for the kinetic equation  a single additional convection term is added to the right hand side of each of the equations (\ref{eq:rhofull}) and (\ref{eq:qfull}), which are $-\nabla\cdot(\mathbf{V_g}\rho)$ and $-\nabla_\alpha(V_{g\alpha}q_\beta)$ respectively. 
With the additional fluxes, the equations can be solved numerically with a step-like gravity field, that is, in a box of length $L$, the gravity takes the values $v_g(x)=g$ for $x\in[-L/2,0)$ and $v_g(x)=-g$ for $x\in[0,L/2]$. With this form, the particles accumulate at the center, while far from $x=0$ the fields decay exponentially, with a sedimentation length,
\begin{align}
    l&=\frac{V^2-2g^2}{4gD_r}+\frac{1}{4gD_r}\sqrt{\left(V^2-2g^2\right)^2+V^2g^2},\\
    &\approx \frac{D_0}{g},
\end{align}
where in the last expression we consider the case $g\ll V$ and we identified the bare diffusivity $D_0$. Note that $l$ is independent of the particle size $\sigma$ because at low densities no collisions take place. Remembering that our model does not consider translational noise, the previous expression is consistent with previous experimental \cite{PhysRevLett.105.088304,PhysRevX.5.011004} and theoretical \cite{Solon2015} findings.

\begin{figure}
    \centering
    \includegraphics[width=\linewidth]{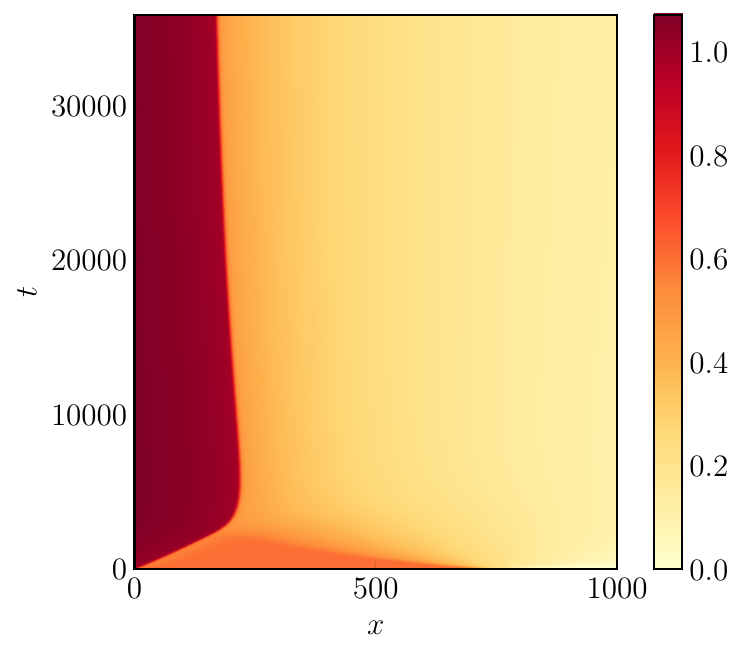}
    \caption{Spatiotemporal evolution of the density $\rho(x,t)$ in the presence of a step-like gravity field, starting from a homogeneous state $\rho(x,0)=0.6$ for $|x|<750$ and zero otherwise, and $q(x,0)=0$, with $D_r=0.01$ and $g=0.05$. The equations were solved in a box with $L=6000$, but only the region up to $x=1000$ is shown. Furthermore, only the section where $v_g(x)=-g$ is presented, as the other side is completely symmetric. Units were fixed by $V=\sigma=1$.}
    \label{fig:gravedadrho}
\end{figure}

\begin{figure}
    \centering
    \includegraphics[width=\linewidth]{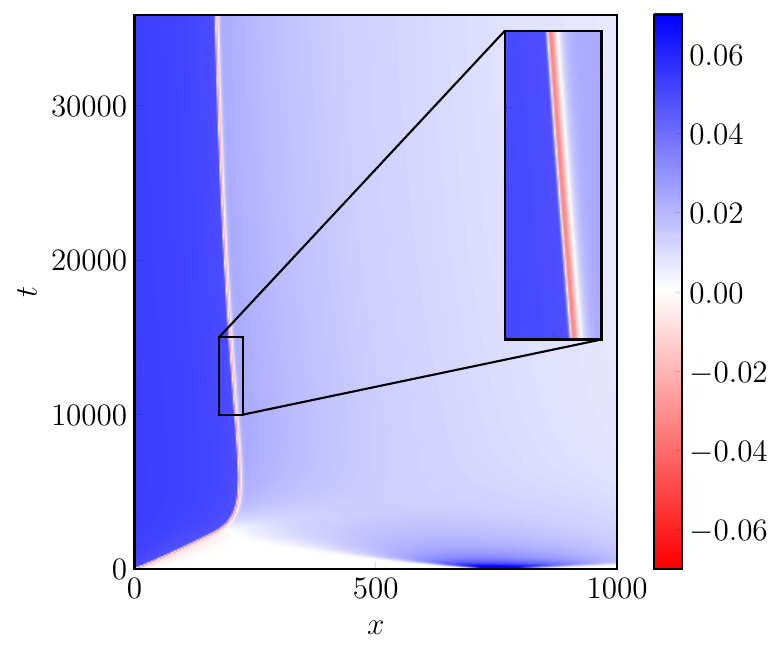}
    \caption{Spatiotemporal evolution of the polarization $q(x,t)$ in the presence of a step-like gravity field, with the same parameters as in Fig.~\ref{fig:gravedadrho}. The inset shows the interface between the dense and dilute regions}
    \label{fig:gravedadq}
\end{figure}

Figures~\ref{fig:gravedadrho} and \ref{fig:gravedadq} show the evolution of the density and the polarization fields in the region where the gravity points to the left, $v_g(x)=-g$. Since the initial homogeneous density is below the critical one, there is no spinodal decomposition. The solution shows how the step-like gravity field effectively mimics a wall at $x=0$, causing the sedimentation of the active particles. The dense region grows non-monotonically towards a stationary state, which may be related to the sedimentation time scale being smaller than the reorientation time. 

It can also be seen that the applied field induces polar order in the particles, causing a polarization against the direction of the gravity, in both the dense and dilute phases. In the dilute region this must be the case for the net flux to vanish in the stationary state. Additionally, for particles to reach that region of space, in average they must be pointing outwards. In the dense region the collisional flux is nonvanishing, but the contributions scale as $\nabla\rho^2$, resulting in small corrections and a positive polarization overall. The exception occurs at the interface, where the polarization is negative. This configuration is needed, because a thin layer of particles pointing towards the center sustain the dense phase. If this was not the case, the net flux of particles would point outward, leading to the eventual disassembly of the cluster.
It should be noted that in order to see the exponential decay in the fields, the solution had to be carried out in a large box. 


\section{Discussion}\label{sec:discussion}
In the high persistence regime for ABP it is possible to build a kinetic theory, which describes particle interactions with an effective collision theory, where they suffer instantaneous effective displacements. This approach has already proven to correctly predict MIPS \cite{PhysRevLett.132.208301} and the effective reduction of the particle speed in dense media~\cite{10.1063/5.0255082}. This kinetic model allows to utilize the Chapman--Enskog method, through which we derive hydrodynamic equations describing the dynamics across multiple time scales. In addition to the particle density, in the high persistence regime, the polarization becomes a slow variable. As a result, closed Navier--Stokes-like equations are obtained, coupling both fields with transport coefficients expressed entirely in terms of the microscopic quantities.

A linear stability analysis shows that the equations derived predict MIPS for a wide range of Péclet numbers, but the theory breaks down for sufficiently high densities. This is not a surprising result, since the equations were derived for ABP gases, where particle correlations were neglected in the kinetic equation. Numerical solutions show that the equations alone fail to accurately describe the full dynamics of the phase separation, requiring the addition of a regularizing term introduced phenomenologically. Another aspect to consider is that the saturation density $\rho_l$, that is, where the effective velocity vanishes, is greater than the close packing one.

Similar methods as those presented in the manuscript can be applied to  derive the hydrodynamic equations for other models of active matter, for example binary mixtures, particles presenting aligning torques, or non-reciprocal interactions. Being able to derive the hydrodynamic equations from the microscopic model gives the possibility to identify the origin of different phenomena or tune the microscopic parameters to achieve desired effects. 

The analysis of the kinetics of phase separation has been the subject an intense research, with a debate if the temporal growth of the domains follows the equilibrium exponents~\cite{thompson2011lattice,PhysRevLett.110.055701,stenhammar2014phase,Gonnella2015-iv,van2019interrupted,caporusso2020motility,caporusso2023dynamics}. In our case, the solution of the hydrodynamic equations in one dimension showed that the coarsening process present the same scaling as in equilibrium. Numerical or analytical studies of the equations  we have derived, with and without noise and in more dimensions, could provide insight into this debate an on the non-equilibrium character of the transition and of active baths in general~\cite{fodor2016far,al2025cold}. Also, the the hydrodynamic equations, where polarization is a relevant field, can help in the analysis of the microphase and macrophase separation, where patches of different orientation appear~\cite{van2019interrupted,caporusso2020motility,caporusso2023dynamics}.

The equations derived here are expected also to be used in the study of active wetting and interfacial dynamics, where the systems are highly polarized. Examples of relevant phenomena include capillary rising~\cite{wysocki2020capillary,fins2024steer}, the growing dynamics~\cite{rojas2023wetting}, the emergent interfacial tension~\cite{fausti2021capillary}, active capillary waves~\cite{turci2021wetting,adkins2022dynamics}, and wetting transitions~\cite{sepulveda2017wetting,turci2021wetting,grodzinski2025hydrodynamic}.

Including particle correlations in the kinetic equation could potentially provide the necessary saturation for (\ref{eq:rhofull}) and (\ref{eq:qfull}) to describe the phase separation dynamics accurately, and give a value of $\rho_l$ smaller than  close packing.
While previous work \cite{PhysRevLett.132.208301} used the pair correlation function for elastic disks to estimate the spinodal density, it was later shown in  Ref.~\cite{10.1063/5.0255082} that this approximation overestimates long-term diffusivity of a tracer ABP for high enough densities. Therefore, a complete theory will require the pair correlation function at contact for precollisional states specific to ABP.

Finally, particle interactions were modeled in the limit of infinite persistence lengths, where directors remain unchanged during collisions. This implies that for a more accurate theory it becomes necessary to develop corrections for large but finite persistence lengths, where due to rotational diffusion, particles detach with new directors. 
 Overall, accounting for spatial correlations and finite persistence corrections is expected to improve the qualitative and quantitative predictions of this theory. 


\acknowledgments
This research was supported by the Fondecyt Grant No.~1220536 and Millennium Science Initiative Program NCN19\_170 of ANID, Chile. M.P.-G. was funded by EPEC-FCFM Doctoral grant of Universidad de Chile.  The authors thank Manuel Mayo and Vicente Garzó for fruitful discussions.

\appendix 
\section{Effective displacements: Derivations and characterization in 2D}\label{app:displ}
Consider two ABP with directors $\mathbf{\hat{n}}_{1,2}$ and a relative vector $\mathbf{r} \equiv \mathbf{r}_2-\mathbf{r}_1=\sigma\hat{\boldsymbol{\sigma}}$ during contact, where $\hat{\boldsymbol{\sigma}}$ is the vector joining both centers. The relative velocity prior to the collision is $V(\mathbf{\hat{n}}_2-\mathbf{\hat{n}}_1)$. Then, the condition for the two particles to collide is that they are approaching, that is, $(\mathbf{\hat{n}}_1-\mathbf{\hat{n}}_2)\cdot\hat{\boldsymbol{\sigma}}_0<0$, with $\hat{\boldsymbol{\sigma}}_0$ the relative vector at the start of the collision.
We define the unit vector $\mathbf{\hat{z}}$ as parallel to $(\mathbf{\hat{n}}_2-\mathbf{\hat{n}}_1)$, that is  $\mathbf{\hat{z}}=(\mathbf{\hat{n}}_2-\mathbf{\hat{n}}_1)/|\mathbf{\hat{n}}_2-\mathbf{\hat{n}}_1|=(\mathbf{\hat{n}}_2-\mathbf{\hat{n}}_1)/\sqrt{2-2\mathbf{\hat{n}}_1\cdot\mathbf{\hat{n}}_2}$, and we define $\theta_0$ as the  angle between $\hat{\boldsymbol{\sigma}}_0$ and $\mathbf{\hat{z}}$, that is, $\cos\theta_0=\hat{\boldsymbol{\sigma}}_0\cdot\mathbf{\hat{z}}$. Note that the collision condition implies that $\rm{cos}\:\theta_0<0$. Similarly, we define the instantaneous angle $\theta$ such that $\cos\theta=\hat{\boldsymbol{\sigma}}\cdot\mathbf{\hat{z}}$.

During contact, the equation of motion of the two particles are
\begin{align}
\dot{\mathbf{r}}_1 &= V \mathbf{\hat{n}}_1 - N\hat{\boldsymbol{
\sigma}}, \label{eq.r1coll}\\
\dot{\mathbf{r}}_2 &= V \mathbf{\hat{n}}_2 + N\hat{\boldsymbol{\sigma}}\label{eq.r2coll},
\end{align}
where $N$ is the normal that prevents the particles from penetrating. The normal can be calculated imposing $(\dot{\mathbf{r}}_2-\dot{\mathbf{r}}_1) \cdot\hat{\boldsymbol{\sigma}}=0$, resulting in $N=V(\mathbf{\hat{n}}_1- \mathbf{\hat{n}}_2)\cdot\hat{\boldsymbol{\sigma}}/2$. The normal vanishes for $\theta=\{\pi/2,3\pi/2\}$, which corresponds to angles of detachment

Adding \eqref{eq.r1coll} and \eqref{eq.r2coll}, gives for the center of mass $\mathbf{R}\equiv (\mathbf{r}_1+\mathbf{r}_2)/2$,
\begin{align}
\dot{\mathbf{R}} &= V (\hat{\mathbf{n}}_1 + \hat{\mathbf{n}}_2)/2. \label{eq.Rcoll}
\end{align}

For the relative vector, as the radial component is fixed by the normal, the only relevant components are the tangential ones. We define the unit vector $\hat{\bm{\theta}}$ to be in the same plane as $\hat{\mathbf{z}}$ and $\hat{\bm{\sigma}}$, but perpendicular to $\hat{\bm{\sigma}}$. We complete the set of spherical unit vectors with $\hat{\bm{\phi}}=\hat{\bm{\sigma}}\times\hat{\bm{\theta}}$. Then, the tangential components for $\dot{\mathbf{r}}=\dot{\mathbf{r}}_2-\dot{\mathbf{r}}_1$ are
\begin{align}
\dot{\mathbf{r}}\cdot\hat{\bm{\phi}} &= V(\hat{\mathbf{n}}_2-\hat{\mathbf{n}}_1)\cdot\hat{\bm{\phi}}\\
&= V|\hat{\mathbf{n}}_2-\hat{\mathbf{n}}_1|\hat{\mathbf{z}}\cdot\hat{\bm{\phi}}\\
&=0
\end{align} 
and 
\begin{align}
\dot{\mathbf{r}}\cdot\hat{\bm{\theta}} &= V(\hat{\mathbf{n}}_2-\hat{\mathbf{n}}_1)\cdot\hat{\bm{\theta}}\\
&= V|\hat{\mathbf{n}}_2-\hat{\mathbf{n}}_1|\hat{\mathbf{z}}\cdot\hat{\bm{\theta}}\\
\sigma\dot\theta&=-V|\hat{\mathbf{n}}_2-\hat{\mathbf{n}}_1|\sin\theta.
\end{align} 
That is, the relative unit vector remains in the $\hat{\mathbf{z}}$-$\hat{\bm{\sigma}}_0$ plane, with the angle evolving according to
\begin{align}
\dot\theta&=-V|\hat{\mathbf{n}}_2-\hat{\mathbf{n}}_1|\sin\theta/\sigma.
\end{align} 
This differential equation must be integrated from the initial condition $\theta_0$ to the moment of detachment at $\pi/2$. That gives the total collision time
\begin{align}
\Delta t^\text{col} &= -\frac{\sigma}{V|\hat{\mathbf{n}}_2-\hat{\mathbf{n}}_1|} \int_{\theta_0}^{\pi/2} \frac{d\theta}{\sin\theta}\\
&= \frac{\sigma}{V|\hat{\mathbf{n}}_2-\hat{\mathbf{n}}_1|} \log\tan(\theta_0/2). \label{sm.dtcol}
\end{align}

At the end of the collision, the relative vector, characterized by having $\theta=\pi/2$ is given by 
\begin{align}
\hat{\mathbf{\sigma}}_\text{end} &= \frac{\hat{\bm{\sigma}}_0 - (\hat{\bm{\sigma}}_0\cdot\hat{\mathbf{z}})\hat{\mathbf{z}}}
{|\hat{\bm{\sigma}}_0 - (\hat{\bm{\sigma}}_0\cdot\hat{\mathbf{z}})\hat{\mathbf{z}}|}\\
&=\frac{\hat{\bm{\sigma}}_0 - \cos\theta_0\hat{\mathbf{z}}}{\sin\theta_0}.
\end{align}
Hence, during the collision, the relative vector has experienced a total displacement
\begin{align}
\Delta\mathbf{r}^\text{col} = \sigma(\hat{\bm{\sigma}}_\text{end} -\hat{\bm{\sigma}}_0).
\end{align}
At the same time, during the collision,  the center of mass has displaced
\begin{align}
\Delta\mathbf{R}^\text{col} = V (\hat{\mathbf{n}}_1 + \hat{\mathbf{n}}_2)\Delta t^\text{col} /2,
\end{align}
where we used Eq.\ \eqref{eq.Rcoll}.
With this, we obtain the total travelled distance  by the particles during the collision
\begin{align}
\Delta\mathbf{r}_1^\text{col} &= \Delta\mathbf{R}^\text{col} - \Delta\mathbf{r}^\text{col}/2,\\
\Delta\mathbf{r}_2^\text{col} &= \Delta\mathbf{R}^\text{col} + \Delta\mathbf{r}^\text{col}/2.
\end{align}

Now, we are in condition to write down the effective displacements $\mathbf{\Delta}_{1/2}$ as the total traveled distance during the collision $\Delta\mathbf{r}_{1/2}^\text{col}$, minus  what they would have traveled during the same time as if there were have been no collision $V\hat{\mathbf{n}}_{1/2}\Delta t^\text{col}$,
\begin{align}
\mathbf{\Delta}_1 &= -\sigma\frac{\hat{\bm{\sigma}}_\text{end}-\hat{\bm{\sigma}}_0}{2}
- V\Delta t^\text{col} \frac{\hat{\mathbf{n}}_1-\hat{\mathbf{n}}_2}{2},\\
\mathbf{\Delta}_2 &= -\mathbf{\Delta}_1. 
\end{align}

Recalling the expressions for $\Delta t^\text{col}$, $\hat{\bm{\sigma}}_\text{end}$, $\hat{\mathbf{z}}$, and $\theta_0$, we notice that effective displacements depend only on $\hat{\mathbf{n}}_2-\hat{\mathbf{n}}_1$ and $\hat{\bm{\sigma_0}}\cdot(\hat{\mathbf{n}}_2-\hat{\mathbf{n}}_1)$. We also note that, by using \eqref{sm.dtcol}, the effective displacements do not depend on $V$.

In three dimensions, the instantaneous unit relative vector $\hat{\mathbf{\sigma}}$ is parametrized by $\theta\in[0,\pi]$, and the azimuthal angle $\varphi\in[0,2\pi]$. In two dimensions, there is no azimuthal angle and now $\theta$ lies in the full range $[0,2\pi]$. Then, two situations can can take place: if $\pi/2\leq\theta_0<\pi$, at the end of the collision $\theta_\text{end}=\pi/2$, but if $\pi<\theta_0\leq3\pi/2$, at the end of the collision $\theta_\text{end}=3\pi/2$. Consistently, Eq.~\eqref{sm.dtcol} has to be modified for the extended range of $\theta$ to  
\begin{align}
\Delta t^\text{col}= \frac{\sigma}{V|\hat{\mathbf{n}}_2-\hat{\mathbf{n}}_1|} \log|\tan(\theta_0/2)|.
\end{align}

Writing the directors as $\hat{\mathbf{n}}_i=(\cos\phi_i,\sin\phi_i)$ and defining the rotation matrix
\begin{align}
R(\alpha)=\begin{pmatrix}
\cos\alpha & -\sin\alpha\\
\sin\alpha & \cos\alpha
\end{pmatrix},
\end{align}
we have
\begin{align}
\hat{\mathbf{z}} &= \left(\frac{\cos\phi_2-\cos\phi_1}{\sqrt{2-2\cos(\phi_2-\phi_1)}}, \frac{\sin\phi_2-\sin\phi_1}{\sqrt{2-2\cos(\phi_2-\phi_1)}}\right),\\
\hat{\bm{\sigma}}_0&=R(-\theta) \hat{\mathbf{z}},\\
\hat{\bm{\sigma}}_\text{end}&=\begin{cases}
R(-\pi/2) \hat{\mathbf{z}}, & \pi/2\leq\theta_0<\pi\\
R(\pi/2) \hat{\mathbf{z}}, & \pi<\theta_0\leq3\pi/2
\end{cases}.
\end{align}

\begin{widetext}
\section{Integral terms of the Chapman--Enskog solution}\label{app:integrals}
At first order in gradients, we need to calculate $J^{(1)}[f^{(0)},f^{(0)}]$, that is 
\begin{equation}
    J^{(1)}[f^{(0)},f^{(0)}]=-\frac{\nabla_\alpha}{4\pi^2}\int \Delta_{1\alpha} \left(\rho + 2q_\beta n_{1\beta}\right)\left(\rho + 2q_\gamma n_{2\gamma}\right)|V\sigma(\mathbf{\hat{n}}_2-\mathbf{\hat{n}}_1)\cdot\hat{\bm{\sigma}}|\Theta[-(\mathbf{\hat{n}}_2-\mathbf{\hat{n}}_1)\cdot\hat{\bm{\sigma}}]d\mathbf{\hat{n}}_2 d\hat{\bm{\sigma}}.
\end{equation}
By expanding the product, the integral can be separated into three terms which have tensors of increasing orders
\begin{equation}\label{eq:J1f0f0}
  J^{(1)}[f^{(0)},f^{(0)}] = -\frac{\nabla_\alpha}{4\pi^2}\left[\rho^2\mathcal{I}_{1\alpha} + 2\rho q_\beta\mathcal{I}_{2\alpha\beta} + 4q_\beta q_\gamma \mathcal{I}_{3\alpha\beta\gamma}\right],
\end{equation}
with 
\begin{align}
     &\mathcal{I}_{1\alpha} =V\sigma \int \Delta_{1\alpha}\lvert(\mathbf{\hat{n}}_2-\mathbf{\hat{n}}_1)\cdot\hat{\bm{\sigma}}\rvert\Theta[(\mathbf{\hat{n}}_1-\mathbf{\hat{n}}_2)\cdot\hat{\bm{\sigma}}]d\mathbf{\hat{n}}_2d\hat{\bm{\sigma}},\\
     &\mathcal{I}_{2\alpha\beta} =V\sigma \int \Delta_{1\alpha}\left(n_{1\beta} + n_{2\beta}\right)\lvert(\mathbf{\hat{n}}_2-\mathbf{\hat{n}}_1)\cdot\hat{\bm{\sigma}}\rvert\Theta[(\mathbf{\hat{n}}_1-\mathbf{\hat{n}}_2)\cdot\hat{\bm{\sigma}}]d\mathbf{\hat{n}}_2d\hat{\bm{\sigma}},\\
     &\mathcal{I}_{3\alpha\beta\gamma}=V\sigma\int \Delta_{1\alpha} n_{1\beta} n_{2\gamma}\lvert(\mathbf{\hat{n}}_2-\mathbf{\hat{n}}_1)\cdot\hat{\bm{\sigma}}\rvert\Theta[(\mathbf{\hat{n}}_1-\mathbf{\hat{n}}_2)\cdot\hat{\bm{\sigma}}]d\mathbf{\hat{n}}_2d\hat{\bm{\sigma}}.
\end{align}
The characterization of the displacements shown in appendix \ref{app:displ} permits direct calculation of each tensor with {\it Mathemtica}. Afterwards, inspection of the result shows 
\begin{align}
    \mathcal{I}_{1\alpha} = -\frac{\pi^2 V\sigma^2}{2}n_{1\alpha},\quad\mathcal{I}_{2\alpha\beta}=\frac{\pi^2V\sigma^2}{4}\left(\delta_{\alpha\beta}-2n_{1\alpha}n_{1\beta}\right),\quad\mathcal{I}_{3\alpha\beta\gamma} = \frac{\pi^2V\sigma^2}{4}n_{1\beta}\delta_{\alpha\gamma},
\end{align}
with $\delta_{\alpha\beta}$ the identity tensor. Substituting these results in Eq.~(\ref{eq:J1f0f0}) yields
\begin{equation}
    J^{(1)}[f^{(0)},f^{(0)}]=-\frac{V\sigma^2\nabla_\alpha}{8}(\rho q_\alpha-\rho^2n_{1\alpha}+2q_\alpha q_\beta n_{1\beta}-2\rho q_\beta n_{1\alpha}n_{1\beta}).
\end{equation}

At order $\epsilon^2$ we need to calculate the moments of $J^{(1)}[f^{(0)},f^{(1)}]$, $J^{(1)}[f^{(1)},f^{(0)}]$ and $J^{(2)}[f^{(0)},f^{(0)}]$. To illustrate, we can take the first moment
    \begin{equation}\label{eq:momenteps2}
        I^{(1)}[f^{(0)},f^{(1)}]=-\frac{\nabla_\alpha}{16\pi^2 D_r}\int \Delta_{1\alpha}(\rho+2q_\beta n_{1\beta})\nabla_\mu[V_\text{eff}(\rho)q_\mu-2V_\text{eff}(\rho)q_\gamma n_{2\mu}n_{2\gamma}]\lvert V\sigma(\mathbf{\hat{n}}_2-\mathbf{\hat{n}}_1)\cdot\hat{\bm{\sigma}}\rvert\Theta[(\mathbf{\hat{n}}_1-\mathbf{\hat{n}}_2)\cdot\hat{\bm{\sigma}}]d\mathbf{\hat{n}}_1d\mathbf{\hat{n}}_2d\hat{\bm{\sigma}}.
    \end{equation}
By expanding the product, we will be left with four integrals to calculate. This time, since we are integrating over both unit vectors, the results of the integrals must be isotropic tensors. Calculating with {\it Mathematica} and direct inspection shows that only two of the terms in Eq.~(\ref{eq:momenteps2}) are nonvaninshing
\begin{align}
     &\int \Delta_{1\alpha}n_{1\beta}\lvert V\sigma(\mathbf{\hat{n}}_2-\mathbf{\hat{n}}_1)\cdot\hat{\bm{\sigma}}\rvert\Theta[(\mathbf{\hat{n}}_1-\mathbf{\hat{n}}_2)\cdot\hat{\bm{\sigma}}]d\mathbf{\hat{n}}_1d\mathbf{\hat{n}}_2d\hat{\bm{\sigma}}=-\frac{\pi^3V\sigma^2}{2}\delta_{\alpha\beta},\\
     &\int \Delta_{1\alpha}n_{1\beta}n_{2\mu}n_{2\gamma}\lvert V\sigma(\mathbf{\hat{n}}_2-\mathbf{\hat{n}}_1)\cdot\hat{\bm{\sigma}}\rvert\Theta[(\mathbf{\hat{n}}_1-\mathbf{\hat{n}}_2)\cdot\hat{\bm{\sigma}}] d\mathbf{\hat{n}}_1d\mathbf{\hat{n}}_2d\hat{\bm{\sigma}}=-\frac{\pi^3V\sigma^2}{4}\delta_{\alpha\beta}\delta_{\gamma\mu}.
\end{align}
With these results, making the contractions yields  $I^{(1)}[f^{(0)},f^{(1)}]=0$. The calculations of the other moments follow analogous procedures and they result in:
\begin{align}
     &I^{(1)}[f^{(1)},f^{(0)}]= 0, \qquad I_{\mathbf{\hat{n}}_1}^{(1)}[f^{(0)},f^{(1)}]=\mathbf{0},\\
     &I_{\mathbf{\hat{n}}_1\nu}^{(1)}[f^{(1)},f^{(0)}]= -\frac{V\pi\sigma^2}{64D_r}\left\{ 2\nabla_\alpha\left[ \rho\overline{\nabla_\alpha \bigl(V_\text{eff}(\rho)q_\nu}\bigr)
 \right]-\nabla_\nu\left[\rho\nabla_\alpha\bigl(V_\text{eff}(\rho)q_\alpha\bigr)\right] \right\},\\
 &I^{(2)}[f^{(0)},f^{(0)}]=\frac{2V\sigma^3}{3\pi}\left[3C\nabla_\alpha^2\rho^2-(1-C)\nabla_\alpha^2q_\beta^2-2(3C-1)\nabla_\alpha\nabla_\beta(q_\alpha q_\beta)  \right]\\
 &I_{\mathbf{\hat{n}}_1\nu}^{(2)}[f^{(0)},f^{(0)}] = \frac{4V\sigma^3}    {9\pi}\left\{(3C-1)\left[\nabla_\alpha^2(\rho q_\nu)-2\nabla_\nu\nabla_\alpha(\rho q_\alpha)+\nabla_\alpha(q_\alpha\nabla_\nu\rho)+\nabla_\nu(q_\alpha\nabla_\alpha\rho)\right]+(3C+1)\nabla_\alpha(q_\nu\nabla_\alpha\rho)\right\}.
\end{align}
In the last integral we used that $\nabla_\alpha(\rho\nabla_\alpha q_\lambda)=\nabla_\alpha^2(\rho q_\lambda)-\nabla_\alpha(q_\lambda\nabla_\alpha\rho)$ to simplify some terms.
    
\end{widetext}

\section{Simulation details}\label{app:simulations}
Simulations are performed with a custom code of active Brownian particles in two dimensions, with all particles having the same diameter $\sigma$ and self-propulsion speed $V$. Particles interact with a WCA potential of intensity $\epsilon=1.0\sigma V$, which is rather soft that permits us to reach long simulation times using the time step $dt=10^{-3}$.  The particles move in a rectangular box with periodic boundary conditions in both directions. The equations of motion are integrated using the Euler-Mayurama method. Finally, the initial condition is with particles placed in a regular square lattice filling all the box to avoid any initial density mode of long wavelength, which would take a very long time to decay and could contaminate the measurements of the intermediate scattering functions. The initial orientations of the particles are random and uncorrelated.

\bibliography{apssamp}

\end{document}